\documentstyle[floats,prl,aps]{revtex} 

\input epsf
\input hyperbasics

\def\spose#1{\hbox to 0pt{#1\hss}}
\def\simlt{\mathrel{\spose{\lower 3pt\hbox{$\mathchar"218$}}
        \raise 2.0pt\hbox{$\mathchar"13C$}}}
\def\simgt{\mathrel{\spose{\lower 3pt\hbox{$\mathchar"218$}}
     \raise 2.0pt\hbox{$\mathchar"13E$}}}

\begin{document}

\def\degrees{\hbox{${}^\circ$\hskip-3pt .}}
\def\spose#1{\hbox to 0pt{#1\hss}}
\def\simlt{\mathrel{\spose{\lower 3pt\hbox{$\mathchar"218$}}
        \raise 2.0pt\hbox{$\mathchar"13C$}}}

\newcommand{\rhos}{\rho_s}
\newcommand{\ps}{p_s}
\newcommand{\vs}{v_s}
\newcommand{\pis}{\pi_s}
\newcommand{\deln}{\delta^N}
\newcommand{\delsy}{\delta^S}
\newcommand{\vn}{v^N}
\newcommand{\vsy}{v^S}
\newcommand{\dotaa}{\displaystyle{{\dot a \over a}}}
\newcommand{\fract}[2]{\displaystyle{{#1 \over #2}}}
\newcommand{\vertsp}{\vphantom{\dotaa}}

\newcommand{\eal}{\! & = & \! }
\newcommand{\eapp}{\! & \approx & \!}
\newcommand{\grad}{\nabla}

\title{DISTINGUISHING CAUSAL SEEDS FROM INFLATION}
\author{Wayne Hu,${}^1$ 
        David N. Spergel${}^{2,3}$ \& 
        Martin White${}^4$\\
}
\address{${}^1$Institute for Advanced Study, School of Natural Sciences\\
Princeton, NJ 08540 \\${}^2$Princeton University Observatory\\
Princeton, NJ 08544 \\
${}^3$Department of Astronomy, University of Maryland\\
College Park, MD 20742 \\
${}^4$Enrico Fermi Institute, University of Chicago\\
Chicago, IL 60637}

\maketitle

\begin{abstract}
\\
\noindent Causal seed models, such as cosmological defects, generically
predict a distinctly different structure to the CMB power spectrum than
inflation, due to the behavior of the perturbations outside the horizon.
We provide a general analysis of their causal generation from isocurvature
initial conditions by analyzing the role of stress perturbations and
conservation laws in the causal evolution.  
Causal stress perturbations tend to generate an isocurvature pattern of
peak heights in the CMB spectrum and shift the first compression,
i.e.~main peak, to
smaller angular scales than in the inflationary case, unless the pressure
and anisotropic stress fluctuations balance in such a way as to reverse
the sense of gravitational interactions while also maintaining
constant gravitational potentials.  
Aside from this case, these causal seed models can be cleanly
distinguished from
inflation by CMB experiments currently underway.
\end{abstract}
\hskip 0.5truecm

\tableofcontents
\section{Introduction}
\label{sec-introduction}

It is now widely recognized that features in the power spectrum of
Cosmic Microwave Background (CMB) anisotropies can be a gold mine of
information for cosmology.
A great deal of experimental effort is being expended in order to map
the CMB accurately over a wide range of angular scales from the ground,
balloons and eventually space.  In addition to providing valuable 
information
about the cosmological parameters, it is becoming clear that the CMB can
teach us much about how the fluctuations were generated in the early universe.
For example in \cite{ourpaper}, it was claimed that by studying the acoustic
signature of the anisotropy spectrum one can test the inflationary paradigm
for fluctuation generation (see \cite{Lid} and references therein for other
inflationary tests).

The key idea in differentiating inflation from other models of structure
formation, such as defects \cite{CriTur,Magetal,Duretal}, is the behavior
of the gravitational potential fluctuations outside the horizon.
In inflation, these potentials are approximately constant while in a viable
defect model, or indeed any isocurvature model, they start out vanishingly
small and are generated as a mode enters the horizon.
Coupled with the effects of photon backreaction, this distinction implies a
different structure in the anisotropy spectrum on small angular scales,
allowing for a test of the inflationary paradigm.
Specifically it was claimed that, with some exotic exceptions, 
isocurvature models
produced spectra whose peaks were phase
shifted with respect to the inflationary models \cite{ourpaper}.  
In a very rough sense, the inflationary driving force excites a cosine mode
whereas the isocurvature one excites a sine mode.
Even if the phase shift were closer to $\pi$ rather than $\pi/2$ radians
\cite{Magetal}, causing the peaks to line up with the inflationary model
once again, the non-monotonic modulation of the peak heights by baryon
drag would allow the defect and inflationary spectra to be distinguished.
We refer the reader to \cite{ourpaper} for more details.

In this paper, we specialize the discussion to causal {\it scaling} models by
applying Turok's \cite{turokletter} mode expansion techniques to the
underlying stress perturbations.  These fluctuations are the fundamental
source of gravitational instability in any isocurvature model
\cite{Bar,KodSas}.  
Detailed discussions of stress perturbations, conservation laws, and gauge in
relativistic perturbation theory as well as their role in causality arguments
are given in the Appendices \ref{sec-perturbation} and 
\ref{sec-causal} respectively.
We explicitly enforce energy-momentum conservation and thus self-consistently
include the response and backreaction of the photon-baryon fluid to the
gravitational sources \cite{ourpaper}.
We show that except for one special case, 
the resultant CMB spectra are easily distinguished
from their inflationary counterpart.  If the 
dynamical effects of isotropic and anisotropic stress are exactly balanced,
a novel situation may arise in which the sense of gravity is reversed 
and hence also the predictions for the acoustic
features in the CMB.  We discuss in detail the model of Turok \cite{turoknew}
which utilizes this mechanism in Appendix \ref{sec-mimic}. 
Thus out of the general class of causal models with
scaling properties only this one case may be confused with inflation from
its acoustic signature.

\section{Conservation Laws and Stress Perturbations}
\label{sec-conservation}

Let us assume that the fluctuations which eventually form large scale
structure in the universe are generated causally from an initially
homogeneous and isotropic Friedman-Robertson-Walker universe.  
Causality, together with energy and momentum conservation, places strong
constraints on the manner in which this can occur.  
Heuristically, energy conservation implies that changes in the energy density
at any location arise only by its ``flow'' across surfaces.
In general, these flows must obey momentum conservation and hence only arise
from stress variations in the matter, e.g.~for a perfect fluid from gradients
in the pressure.
It instructive first to consider the simple case of a non-relativistic fluid.
We shall then show how these arguments manifest themselves in
relativistic perturbation theory with a more general stress-energy 
tensor for the matter. 

\subsection{Non-relativistic Example}
\label{ss-example}

Here we first examine the evolution of perturbations in a simple
non-relativistic fluid, perhaps with viscosity, but ignoring gravitational
effects. Energy-momentum conservation for the perturbations is described by
the linearized continuity and Euler fluid equations
\begin{equation}
\begin{array}{rcl}
\dot \delta \eal -\partial_i v_i, \\
\rho\dot v_i \eal \partial_i p - \partial_j \Pi_{ij}, \\
\end{array}
\label{eqn:simplefluid}
\end{equation}
where summation is implicit, $\delta = \delta\rho/\rho$ is the density
fluctuation, $v_i$ is the bulk velocity, $p$ is the pressure, and $\Pi_{ij}$
is the viscous or anisotropic stress tensor.
Density fluctuations can only be generated by fluid flows.  A Fourier
decomposition of the perturbation implies that compared with velocities,
the density must be suppressed by a factor of $k$ at long wavelengths.
However momentum conservation constrains the form of such flows:
they cannot be present initially and thus must be generated by pressure
gradients.  The Fourier decomposition shows that velocities should be
suppressed with respect to pressure fluctuations by a factor of $k$ at
long wavelengths.  Hence density fluctuations generically scale as $k^2$
times the pressure fluctuations in a fluid or $k^4$ in the power spectrum.
This is the familiar result that causal flows of matter will establish a
$k^4$ density spectrum even when no density perturbations exist initially
\cite{Zel,Pee74,CarSil,RobWan}.

The Poisson equation implies that the resultant potential fluctuations scale
as the pressure itself.  In a relativistic context, potential fluctuations are
equivalent to curvature fluctuations in the spatial metric.  The fact that
the generator of density and curvature fluctuations is causal requires that
initially they must vanish.  Hence we refer to such models for fluctuation
generation as {\it isocurvature} models.

The form of the pressure perturbation itself is not arbitrary.  In fact, if the
pressure perturbations are adiabatic
$\delta p = (\dot{p}/\dot{\rho})\delta\rho \equiv c_s^2\delta\rho$,
then energy-momentum conservation requires $\delta\rho=0$ and fixes
$\delta p=0$, so that it cannot generate density perturbations.
Thus it is only the non-adiabatic pressure or ``entropy'' perturbation that
can causally produce density fluctuations \cite{Bar}
\begin{equation}
p\Gamma = \delta p - c_s^2 \delta \rho.
\end{equation}
In general, there are many possible sources of non-adiabatic pressure,
but causality constrains their behavior by
requiring that their fluctuations be uncorrelated outside the horizon.
One natural way to obtain them is to assume the fluid is composed of
a sum over $i$ particle constituents. 
In this case,
\begin{equation}
p\Gamma = \sum_i [p_i \Gamma_i + (c_i^2 - c_s^2)\delta\rho_i],
\end{equation}
so that if the sound speed $c_i$ in the components does not equal the total
sound speed, i.e.~the equation of state for the components differ, then the
initial condition $\delta\rho = \sum_i \delta \rho_i = 0$ implies that
non-adiabatic pressure perturbations {\it must} be generated.
In the cosmological setting, concrete examples of this mechanism include the
baryon and axion isocurvature models as well as cosmological defect
scenarios.
This idea, that density fluctuations may be balanced to satisfy
total energy-momentum conservation, is conventionally referred to
as {\it compensation}.  Compensation once established
initially is maintained by energy-momentum conservation; 
there is no need to enforce it by hand as often done in the
literature \cite{Magetal}.

Now let us consider the anisotropic stress.  Internal friction or viscosity is
generated when there is relative motion between various parts of the fluid.
The anisotropic stress tensor thus scales as the spatial derivatives of the
velocity field and to lowest order, the first derivative
(see e.g.~\cite{LanLif}). 
By momentum conservation, 
we know that the velocity field vanishes initially.
Hence anisotropic stress is only generated after pressure gradients set up
bulk motion.  The scaling in $k$-space is that of $k$ times the velocity
fluctuation or $k^2$ times the pressure fluctuations.
The Euler equation thus implies that {\it at large scales} the generation of
bulk velocities and hence density and potential fluctuations through
anisotropic stress is subdominant.

This simple example shows that the energy-momentum conservation equations
automatically build in causal behavior.  The problem of considering the
effects of causality thus reduces to the establishment of causal initial
conditions and the enforcement of energy-momentum conservation as the universe
evolves under the stresses of the matter.  

\subsection{Relativistic Generalization}
\label{ss-relativistic}

Two issues complicate the simple picture of the last section.  
The first is that we must possess a model for how the stress perturbations
evolve.  We shall return to consider causal constraints on their
behavior in the next section.  The second
is that in relativistic perturbation theory,  
the stress-energy tensor of the
matter is {\it covariantly} conserved.  Hence the continuity and 
Euler relations of Eq.~(\ref{eqn:simplefluid}) become
\begin{equation}
T^{0\nu}_{\hphantom{0\nu};\nu}=0, \qquad T^{i\nu}_{\hphantom{i\nu};\nu}=0.
\end{equation}
Because metric terms enter these equations, the form that the causal constraint
takes depends on the metric representation, i.e.~the gauge.
For example, the continuity equation is altered by changes in the spatial
metric.  The simplest example is that of the stretching of space due to the
background expansion, which dilutes the number density of particles in physical
space.  Likewise perturbations to the spatial metric cause similar effects to
the density perturbation.  To disentangle metric effects on the generation of
perturbations from the truly causal evolution by flows, it is desirable to find
a representation of perturbations that obeys an ordinary conservation law.
In this context, two quantities have been often discussed in the literature:
the stress-energy pseudo-tensor $\tau_{\mu\nu}$ \cite{VeeSte,PenSpeTur} 
and the comoving curvature perturbation $\zeta$ \cite{Bar,AbbTra}.

To understand the problem, it is perhaps useful to recall first the issue of
gauge choice. In relativistic perturbation theory, one has the freedom to
choose which spatial surface, and what coordinate system on this surface,
to use in defining the perturbations.
Gauge freedom can be both a complicating annoyance and a very convenient tool,
but poses no real obstacle for applying relativistic perturbation theory. 
Once the initial conditions are properly established, covariant conservation of
the stress-energy tensor properly and causally evolves the fluctuations in any
gauge.  In particular, all gauges will agree on {\it physical observables},
e.g.~CMB anisotropies.  Three gauges choices, for which we give detailed
properties in Appendix \ref{sec-perturbation}, 
are in common use.  Let us briefly note here their
benefits and drawbacks before specializing the discussion to the relativistic
analogue of the initial conditions described in the previous section.

Perhaps the most popular gauge choice is that of synchronous gauge, 
where the perturbations appear only in the space-space part of 
the metric (see e.g.~\cite{Pee}).  
In this gauge, the spatial hypersurfaces on which one defines 
the perturbations are orthogonal to constant-time hypersurfaces 
and proper time corresponds to coordinate time.
Thus this coordinate system is natural for freely-falling observers or cold
dark matter particles.
The drawback of this gauge is that the density perturbations are not 
easily related to the observable anisotropy and the gravitational sector 
is non-intuitive.
One must be careful to compute observables as the individual components of
this gauge can be quite misleading. 
 
The most familiar gauge from courses in relativity is the conformal 
Newtonian gauge.  In this gauge, the metric is diagonal: the 
space-space part gives the curvature perturbation, and the time-time 
part the gravitational potential.  This gauge has been frequently used 
in analytic work on CMB anisotropies because the representation of the 
gravitational (Sachs-Wolfe) effects is simple and the density 
perturbations correspond closely to the CMB anisotropy.  The gauge 
can be difficult to work with numerically, and extreme care must 
be taken with the initial conditions.

For work involving causality, the obvious gauge choice is the comoving
gauge, also known as the total-matter gauge and velocity-orthogonal isotropic
gauge.  This gauge is difficult to conceptualize, since it contains
an off-diagonal time-space perturbation.
However as we shall show in Appendix \ref{sec-causal}, 
the spurious effects of density dilution
(from stretching of the spatial metric) 
which complicate the analysis of the
conservation laws are absent in this gauge.
More specifically, the curvature perturbation $\zeta$ 
in this gauge (superscript $T$) is generated only
by pressure (non-adiabatic if the curvature vanishes initially) and
anisotropic stress fluctuations $\pi$ (see Eq.~(\ref{eqn:zetadot}),
definitions in \S \ref{ss-comoving}, 
and \cite{Wei} for anisotropic stress terms in
the relativistic fluid context)
\begin{equation}
\dot\zeta = -\dotaa {1 \over \rho + p}(\delta p^T - {2 \over 3}\pi),
\end{equation}
where $a$ is the scale factor and temporal derivatives are hereafter
with respect to conformal time $\tau = \int dt/a$.
In the case of a more general stress-energy tensor, we can merely
replace $\delta p$ and $\pi$ by the isotropic and anisotropic
scalar components of $T^i_{\hphantom{i}j}$ 
[see Eq.~(\ref{eqn:stressenergy})].

The direct dependence of the curvature on stress perturbations
implies that the causality argument in this gauge is the most similar
to the non-relativistic case discussed in the previous section.
For example, in the absence of these stresses, {e.g.}~in the inflationary
example with adiabatic fluctuations, the curvature is simply constant outside
the horizon to leading order \cite{BarSteTur,Lyt}.
Thus the proper generalization of the causal argument to the relativistic
context is that the curvature on the comoving hypersurfaces $\zeta$
vanishes initially and is only generated
by the causal motion of matter (see \cite{ourpaper,Bar} and 
Appendix \ref{sec-causal}).

{}From this condition, it is simple to reconstruct the causal constraint
in the two other gauges from gauge transformations (see \S \ref{ss-initial}).
The curvature on Newtonian hypersurfaces is directly proportional
to the density fluctuations on the comoving hypersurfaces [see 
Eq.~(\ref{eqn:poissontmg})].
This suggests that the isocurvature condition for the total matter
gauge is the same as that of the Newtonian gauge.  We show in 
Appendix \ref{sec-causal} 
that this intuition is correct, up to an irrelevant decaying
mode [see Eq.~(\ref{eqn:decaying})], if the equation of state is constant.
For the synchronous
gauge, the condition that $\zeta =0$ is identical to the assumption
that the pseudo-energy $\tau_{00}$ and the pseudo-momentum density
$\tau_{0i}$ defined in Eq.~(\ref{eqn:pseudose}) vanishes initially 
\cite{VeeSte}. 
These are components of the stress-energy pseudo-tensor commonly 
employed in the literature, which likewise obeys an ordinary 
conservation condition
\begin{equation}
\dot \tau_{00} = \partial^i \tau_{0i},
\end{equation}
as one would expect.
Thus these three sets of initial conditions: vanishing of the comoving 
curvature $\zeta$, Newtonian curvature $\Phi$, 
and $\tau_{00}$, $\tau_{0i}$,
are essentially equivalent.  Once these conditions are established,
energy-momentum conservation causally evolves the perturbations under
the influence of spatial stresses, generating properties such as 
a $k^4$ scaling in the power spectrum of the pseudo-energy 
and comoving density perturbation.   Let us now turn to the
question of causal stress evolution.

\subsection{Scaling Stress Sources}
\label{ss-scaling}
 
Causality implies that no measurable quantity, e.g.~the 
fields and stress-energy components, can
have superhorizon scale correlations.  This implies
that their power spectrum behaves 
as ``white noise'', $k^0$ to leading order for $k\tau \ll 1$,
unless other symmetries exist to eliminate even this contribution
(e.g.~energy-momentum conservation and the comoving density 
perturbation, see also \S \ref{ss-stress}).  In 
Appendix \ref{sec-causal},
we show that for models with scalar fields, this constraint
limits the superhorizon scale behavior of
all of the stresses: the isotropic stress $p_s$ behaves 
as $k^0$ and the anisotropic stresses,
which depend only on  spatial derivatives of fields, behaves
as $k^2$ for $k\tau \ll 1$.  

Turok \cite{turokletter} raises the interesting question of what
general statements for the CMB anisotropy spectrum can be made
if one combines causality with the {\it scaling ansatz}.
The scaling ansatz is a powerful tool for analyzing the
dynamics and predictions of defect models \cite{Kib,Vil,Hind}.
It implies that defect networks have only one characteristic scale, set
by the current horizon size.  Thus, for example, a string network
a few thousand years after the big bang has the same correlations
as a string network a nanosecond after the big bang.
This scaling ansatz has been useful for studying the dynamics
of the non-linear sigma model, a simple approximation
to defect dynamics, that has scaling solutions
in both the matter and radiation dominated epochs\cite{Tur,Fil}.

For our purposes, scaling may be defined more phenomenologically as the
assumption that the power (per $\log k$) in the metric fluctuations,
i.e.~in the Newtonian curvature fluctuation $k^3 |\Phi|^2$, and potential
fluctuation $k^3 |\Psi|^2$ are both the same on each scale at horizon crossing
$k\tau=1$ and evolve in a self-similar fashion 
\begin{equation}
\Phi = k^{-3/2} f(k\tau), \qquad \Psi = k^{-3/2} g(k\tau).
\label{eqn:scaling}
\end{equation}
The Newtonian potential and curvature perturbations along with their
evolution are directly related to
the gravitational redshifts experienced by a photon \cite{SacWol}.
Thus {\it any} model which can explain the flatness
of the large angle CMB spectrum seen by {\it COBE} \cite{Benetal} 
must also obey the scaling ansatz 
at least approximately.  The inflationary scenario naturally
generates such fluctuations with $f(k\tau) = g(k\tau) = $ constant.  We
must now seek a causal mechanism for their generation through
stress perturbations.  The ansatz cannot simply be imposed 
on the metric fluctuations since this does not guarantee that a consistent
solution of the conservation and Einstein-Poisson equations 
exist.

Consider first the Newtonian curvature $\Phi$.
{}From the arguments of \S \ref{ss-example} \& \S
\ref{ss-relativistic}, made rigorous by the gauge considerations of
the Appendices \ref{sec-perturbation} and \ref{sec-causal}, 
a pressure fluctuation source $p_s$ generates a comoving gauge
density perturbation (superscript $T$) of order
\begin{equation}
\rho \delta^T \sim (k\tau)^2 p_s,
\end{equation}
and hence from the Newtonian Poisson equation (\ref{eqn:poissontmg})
\begin{equation}
\begin{array}{rcl}
k^2\Phi & \! \sim \! &  4\pi G a^2 \rho \delta^T \\
& \! \sim  \! &  4\pi G (a^2 p_s) (k\tau)^2 .
\end{array}
\end{equation}
For white noise pressure perturbations, 
the scaling ansatz Eq.~(\ref{eqn:scaling}) then requires
\begin{equation}
\begin{array}{rcl}
a^2 p_s  \propto \tau^{-1/2},
\end{array}
\end{equation}
for $k\tau \ll 1$. 
Thus if we adopt this ansatz for the pressure source, energy-momentum
conservation will naturally generate scaling behavior in $\Phi$. 
Note also that white noise pressure perturbations imply
white noise curvature fluctuations.

Turok \cite{turokletter} points out that to study possible behaviors around
and after horizon crossing, we can decompose the source into basis functions
that satisfy scaling and a strict lack of correlations outside the horizon
\begin{equation}
\left\langle a^2 p_s(k,\tau)\ a'^2 p_s(k,\tau') \right\rangle =
 \tau^{-1/2} \tau'^{-1/2} \sum_A \sum_{A'} P_{AA'} f_A(k\tau) f_{A'}(k\tau'),
\end{equation}
for which in real space
\begin{equation}
\left\langle f_A(r,\tau) f_{A'}(0,\tau') \right\rangle = 0
  \qquad {\rm for} \qquad r > \tau + \tau'.
\end{equation}
The symmetry in $k\tau$ implies that a diagonal basis
exists where $P_{AA'}=\delta_{A A'}P_A$ \cite{turokletter}.  
However, for illustrative purposes, we follow Turok in employing 
\begin{equation}
f_A \equiv \fract{\sin(Ak\tau)}{(Ak\tau)},
\label{eqn:basis}
\end{equation}
as a convenient basis, where $0 < A < 1$.
We shall therefore adopt in the next section pressure sources of the form
$a^2 p_s \propto \tau^{-1/2}f_A$ which differs from Turok's suggestion
of $a^2 (\rho_s+3p_s) \propto \tau^{-1/2}f_A$ (see also 
Appendix \ref{sec-mimic}). 
Our 
assumption follows from scaling and causal constraints on stresses
and allows the density evolution to be naturally determined
by energy-momentum conservation from the source stresses. 

Now let us consider the Newtonian gravitational potential 
[see Eq.~(\ref{eqn:poissonnewt})],
\begin{equation}
\Psi = -\Phi - 8\pi G a^2 \pi_s/k^2.
\label{eqn:psipi}
\end{equation}
Thus the scaling ansatz for $\Phi$ holds equally well for $\Psi$ 
except in the presence of anisotropic stress contributions $\pi_s$.  
To produce a flat CMB anisotropy spectrum, any such contributions
must also obey a scaling relation
\begin{equation}
a^2 \pi_s \propto \tau^{-1/2} f_B(k\tau).
\end{equation}
If the universe is isotropic initially, anisotropic stress like the
comoving density perturbation can only be generated by causal motion
of matter
implying a $k^2$ scaling
for $k\tau \ll 1$ (see \S \ref{ss-example} for an example).  
The same arguments employed in deriving the
general form of the causal pressure source imply (see \cite{turoknew}
for an analogous derivation)
\begin{equation}
f_B(k\tau) = {6 \over B_2^2 - B_1^2}
	      \left[ {\sin(B_1 k\tau) \over (B_1 k\tau)}
	      -{\sin(B_2 k\tau) \over (B_2 k\tau)} \right],
\label{eqn:basispi}
\end{equation}
where $0 < (B_1, B_2) < 1$ and
we have normalized the function to behave as $(k\tau)^2$ on
small scales.
Thus Eq.~(\ref{eqn:basis}) and (\ref{eqn:basispi}) represent the 
mode decompositions of a scaling isotropic and anisotropic stress
perturbation which strictly obeys causal constraints for a
lack of correlations above the horizon.

\section{Implications for the CMB}
\label{sec-implications}

\subsection{Acoustic Sources and Signatures}
\label{ss-acoustic}

Let us first review the formalism set up in \cite{ourpaper} for
calculating the acoustic oscillations in the CMB for a model with
external gravitational sources. 
To avoid obscuring the main physical points, we have relegated the
technical details to the Appendices \ref{sec-perturbation}
and \ref{sec-causal}.
The basic idea is that one solves the equations for the fluid and metric
evolution under the influence of sources which are assumed to interact with
the fluids only through gravity.
For CMB studies, it is convenient to choose a Newtonian gauge condition to
represent these effects, since the perturbations are more easily interpreted
in this gauge than in the synchronous or comoving gauge.
The Einstein equations tell us that the matter fields generate Newtonian
metric perturbations: specifically the curvature $\Phi$ and gravitational
potential $\Psi$.

Under the methods of \cite{ourpaper}, the gravitational
contributions of the photon-baryon fluid are separated from
the other sources,
$\Phi = \Phi_{\gamma b} + \Phi_s$ and
$\Psi = \Psi_{\gamma b} + \Psi_s$.
The photon-baryon oscillator equation is then solved in the presence of
$\Phi_s$ and $\Psi_s$.
If we assume that the source is composed of seeds, i.e.~a component
whose stress-energy tensor makes a small perturbation to the background,
its contribution is
[see Eqs.~(\ref{eqn:stressenergy}) and (\ref{eqn:conserve})]
\begin{equation}
\begin{array}{rcl}
(k^2 - 3K)\Phi_s \eal 4\pi G a^2 (\rhos +
	 3{\displaystyle{\dot a \over a}}\vs/k ), \\
k^2 (\Psi_s + \Phi_s) \eal -8\pi G a^2 \pis,
\end{array}
\label{eqn:phis}
\end{equation}
where we have simply labeled the scalar 
terms of the stress-energy tensor
of the seeds in the fluid convention without loss of
generality\footnote{For reference, note that the
relationship between \cite{CriTur,turokletter} and our notation is 
$\Theta_{00} = a^2\rhos$, $\Theta_{ii} = 3a^2\ps$, $\Theta_S = -a^2\pi_s$
and $\Pi = -a^2 k v_s$.} 
[see Eq.~(\ref{eqn:stressenergy})].
Thus since defect seeds merely represent 
a special case of an external
source, they may easily be treated under this formalism.

To summarize the results of \cite{ourpaper}, it was established that
isocurvature initial conditions, in the sense of \S \ref{ss-relativistic} and  Appendix \ref{sec-causal},
robustly predict an anticorrelation between the source curvature
and CMB temperature perturbations at horizon crossing 
during radiation domination.
The underlying reason is obvious from the causal arguments of \S 
\ref{sec-conservation}:
changes in the source energy density must be compensated by an opposing change
in the radiation density before bulk motion has had a chance to redistribute
the matter.  
Since correlations and anticorrelations with the curvature represent
compressions and rarefactions in potential wells respectively under
normal conditions, the acoustic signature can distinguish between these
cases (see \S \ref{ss-assumptions} and Appendix \ref{sec-mimic} 
for exceptions).  
The specific signature is provided by the drag baryons induce on the
photon-baryon fluid which enhances compressions over rarefactions.
Thus the signature of an inflationary model is given by the ratio of the
acoustic peak locations, which measures the phase of the acoustic 
oscillation, and an enhancement of the odd peak
heights.
It was found that though a distinctly different set of ``sine''
peak ratios was
a common prediction of isocurvature models, details of the source evolution
could be tuned to reproduce the inflationary case (see also \cite{Magetal})
so that the peak height test is also necessary.  We now consider whether
additional assumptions, such as scaling in a strictly causal stress model,
can produce further robust distinctions.

\begin{figure}
\begin{center}
\leavevmode
\hskip-0.5truecm
\epsfxsize=3.5in \epsfbox{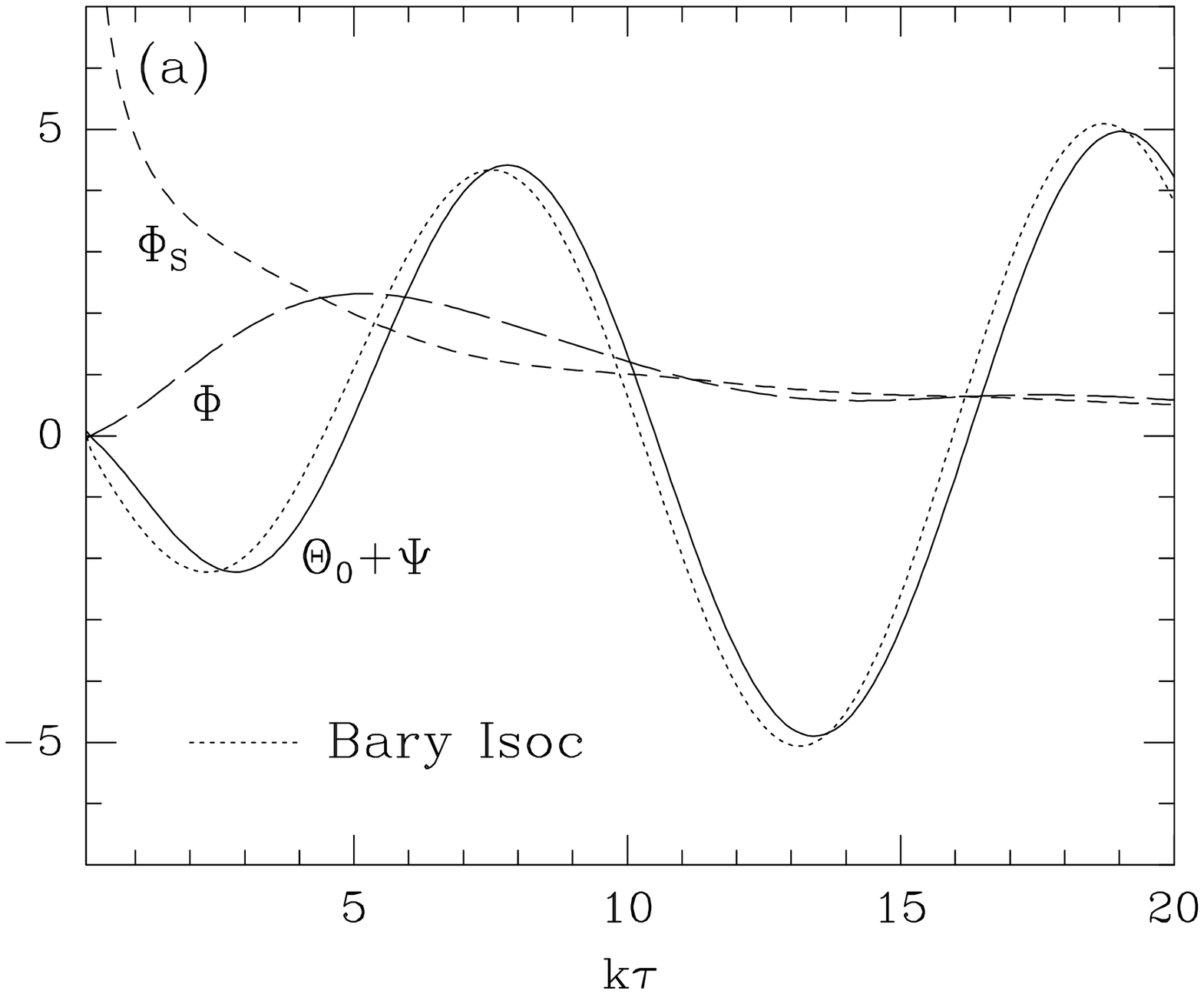} 
\epsfxsize=3.5in \epsfbox{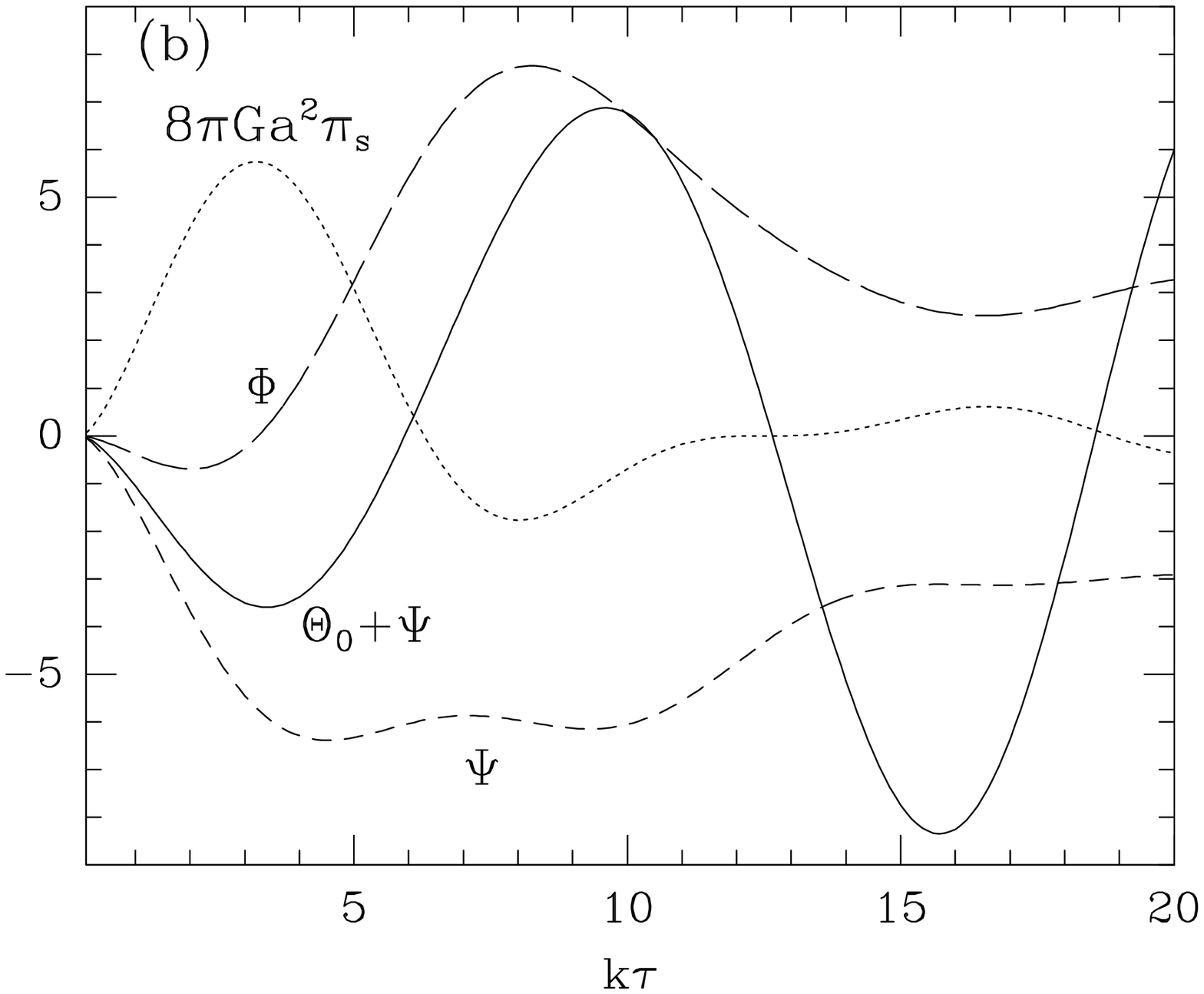} 
\end{center}
\caption{(a) Pressure scaling source. 
The effective temperature $\Theta_0+\Psi$, total curvature
perturbation $\Phi$, and the contribution from the source $\Phi_s$, produced
by $p_s$ assuming Eq.~(18) with $A=1$.  
Notice that the temperature fluctuations are similar to the
canonical prediction of a baryon-isocurvature model (dotted line), 
not inflation.
(b) Anisotropic stress scaling source.  Evolution under $\pi_s$ 
assuming Eq.~(19) with $B_1=1$, $B_2=0.5$.
Photon domination is assumed here and in Figs.~3 and 4.
}
\label{fig:pressure}
\end{figure}

\subsection{Scaling Ansatz for Pressure}
\label{ss-pressure}

Let us now specialize the analysis of \cite{ourpaper} to the case where the
source pressure fluctuations are from seeds 
that obey the scaling ansatz 
discussed in \S \ref{ss-scaling}.  Specifically, let us break the pressure source into
contributions that behave as
\begin{equation}
4\pi G a^2 p_s = 
\tau^{-1/2} f_A = 
\tau^{-1/2} \fract{\sin(Ak\tau)}{(Ak\tau)},
\label{eqn:pressurebasis}
\end{equation}
with $0 < A < 1$.  
This choice is similar to and inspired by the ansatz of 
\cite{turokletter}
but replaces the assumption for $a^2(\rhos+3\ps)$ with the analogous one for
$a^2\ps$ since stress fluctuations are the fundamental source of
causally-seeded perturbations.  
This allows energy-momentum conservation to fix the form of density
perturbations naturally from the stress fluctuations and permits
a wider class of possible seed sources (see Appendix \ref{sec-mimic}).
For simplicity, we here assume that the seed anisotropic stress
$\pi_s=0$ and postpone discussion of its effect until the next 
section.

We show the evolution of the system under the source 
Eq.~(\ref{eqn:pressurebasis}) in Fig.~\ref{fig:pressure}a.  
We have chosen $A=1$ since this is the most extreme of the causal modes in
that it produces features in the source as soon as causally possible.
The initial conditions require a vanishing comoving
curvature $\zeta=0$ or equivalently vanishing stress-energy pseudo-tensor
components $\tau_{00}=0=\tau_{0i}$.  As long as the initial conditions are
set early enough so that the pressure has not generated significant
perturbations, this may be satisfied by setting the individual energy density
and momentum density perturbations of the fluids and sources to zero.
Note that in the more general context in which the source fluctuations are 
directly related to density fluctuations, one must be more careful in setting
up consistent compensated initial conditions.

We make the important assumption here that the universe is radiation
dominated as the fluctuation enters the horizon which we will discuss
further below (see also \cite{ourpaper}). 
Notice that this model follows the predictions of a canonical
isocurvature model $\Phi_s \propto (k\tau)^{-1}$ (e.g.~a baryon
or axion isocurvature model \cite{ourpaper,HuSug}), quite closely.  
Here the effective temperature perturbation
$\Theta_0 + \Psi$ is composed of
the
temperature fluctuation on Newtonian surfaces
$\Theta_0=\delta_\gamma^N/4$
[see Eq.~(\ref{eqn:fluidsynchnewt})],
and the Newtonian potential $\Psi$ which accounts for the gravitational
redshift or Sachs-Wolfe effect.

Note that the sign change in the pressure at 
$k\tau = \pi$ has no direct relevance to the question of acoustic
phase.  The action of a source near or outside the horizon 
generically drives a sine mode acoustic wave due to feedback from
the self-gravity of the photon-baryon fluid. The situation for
which these arguments fail is if feedback is unimportant, i.e.~if 
the universe is {\it fully} matter-dominated when the mode entered
the horizon.  
These issues are treated in much greater detail in \cite{ourpaper}.

\begin{figure}
\begin{center}
\leavevmode
\epsfxsize=4.5in \epsfbox{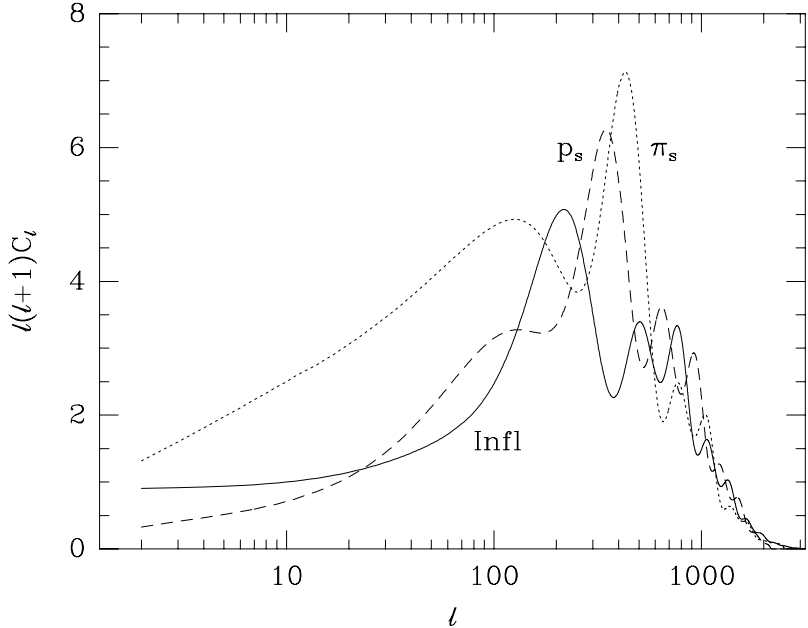} 
\end{center}
\caption{The anisotropy power spectrum, $\ell(\ell+1)C_\ell$, vs multipole
number $\ell\sim\theta^{-1}$.
The solid line is the inflationary prediction.
The dashed line assumes a pressure source with the form of Eq.~(18) for $A=1$.
The dotted line assumes an anisotropic source source with the form of Eq.~(19)
for $B_1=1$ and $B_2=0.5$.
All curves assume the same background cosmology $\Omega_0=1$, $h=0.5$,
$\Omega_b h^2 =0.0125$.  Notice that the predictions are out of phase and
that even rather than odd peaks are prominent in the non-inflationary models.}
\label{fig:cls}
\end{figure}

To demonstrate that this potential loophole is not a concern for reasonable
cosmological parameters, we perform a full calculation of the $A=1$ model by
solution of the complete Boltzmann equations (see e.g. \cite{HSSW}).
Specifically, the model includes cold dark matter and three species of
massless neutrinos with standard recombination and cosmological parameters
of $\Omega_0=1, \Omega_0 h^2 = 0.25, \Omega_b h^2 = 0.0125$.  Even if the
matter-radiation density ratio at last scattering were as high as
$\rho_m/\rho_r \approx 16$, as in this case, the acoustic signature remains
distinctly isocurvature.
In Fig.~\ref{fig:cls}, we compare the power per log $\ell\sim\theta^{-1}$ in
anisotropies $\ell(\ell+1)C_\ell$ for this model with the standard
inflationary model for the same cosmological parameters.
We have explicitly checked by the Boltzmann calculation that in a universe
with the standard thermal history and reasonable matter content
$\Omega_0 h^2 \simlt 1$, the loophole provided by the absence of photon
backreaction does not exist.  

Note that the calculation of Fig.~\ref{fig:cls} also includes gravitational
effects between last scattering and the present, i.e.~the integrated
Sachs-Wolfe (ISW) effect, in addition to the acoustic temperature fluctuation
displayed in Fig.~\ref{fig:pressure}.
It thus corresponds to the {\it physically observable} total anisotropy from
the pressure fluctuations defined by the model.  We comment on the importance
of calculating the observable anisotropy in Appendix \ref{sec-mimic}. 

\begin{figure}
\begin{center}
\leavevmode
\hskip -0.5truecm
\epsfxsize=3.5in \epsfbox{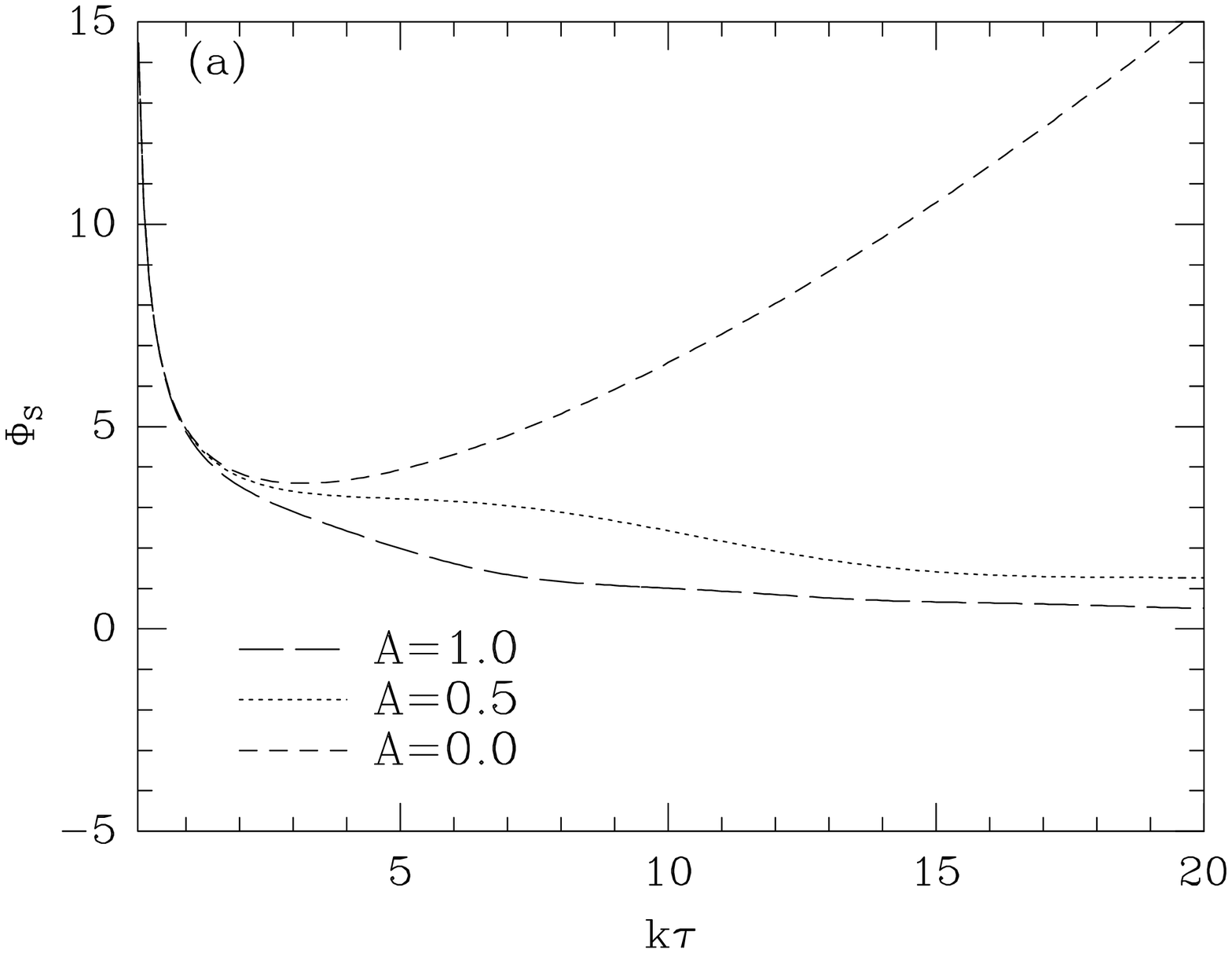} 
\epsfxsize=3.5in \epsfbox{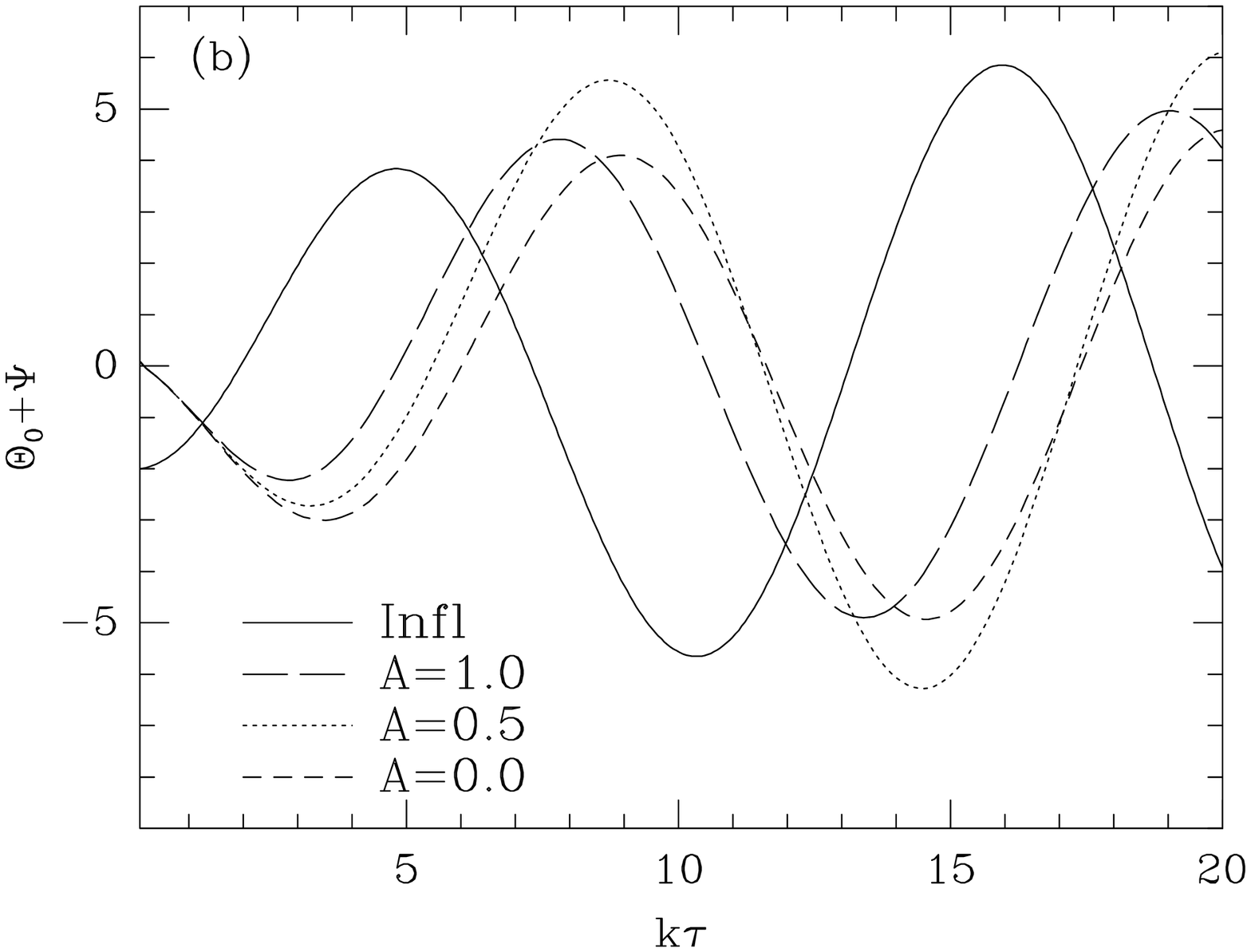} 
\end{center}
\caption{We show (a) the source curvature $\Phi_s$ and (b) the effective
temperature $\Theta_0+\Psi$ for the family of pressure sources of Eq.~(18).  
In all cases, the effective temperature approximately follows the canonical
isocurvature evolution from Fig.~1, which is very different from the
inflationary case (solid line in panel b).}
\label{fig:family}
\end{figure}

It is interesting to consider the full range of $0 < A < 1$ since the modes
may be superimposed to produce a more general case.  
Fig.~\ref{fig:family}a shows that varying $A$ dramatically changes the
behavior of the curvature source inside the horizon ($k\tau \ge 1$).  
The source falls rapidly, has a peak at $k\tau_{\rm peak}\propto A^{-1}$,
then falls again.  On the other hand, when one examines its effect on the
photon-baryon fluid in Fig.~\ref{fig:family}b, the change with $A$ is much
less dramatic.  
In fact for all $0 < A < 1$, the acoustic oscillations follow the
isocurvature pattern of a near sine mode with even compressional (positive)
peaks.  This result is also easy to understand from \cite{ourpaper}.
It is far easier to stimulate an acoustic mode as the fluctuation crosses the
Jeans scale (i.e.~the sound horizon) than after it.  Even the dramatic changes
in the curvature inside the horizon shown in Fig.~\ref{fig:family} are not
sufficient to overwhelm the signal created at crossing.
These arguments apply to any slowly-varying source of metric fluctuations
inside the horizon.  
It is possible to obtain somewhat 
more rapidly-varying features by interfering
different $A$ modes.  
This may occur if additional symmetries in the source eliminate
the white noise pressure contributions (see Appendix \ref{ss-stress}) and 
produces effects similar to anisotropic
stress sources considered in the next section [cf. Eqs.~(\ref{eqn:basis})
and (\ref{eqn:basispi})].
One should also be careful in that although this
acoustic signature is robust for individual $A$ modes, 
it may be difficult to observe in the low
$A \simlt 0.3$ case.  Since after last scattering potential fluctuations
eventually decay, additional contributions from the ISW effect 
can mask this signal.  They do not however possess oscillatory features 
and hence cannot be employed to mimic the inflationary spectrum.

Thus we come to the conclusion that this whole class of
pressure scaling models produces an acoustic signature
that bears the canonical isocurvature stamp: a sine mode oscillation
with a rarefaction-compression-rarefaction pattern that leads to 
even peak enhancement from the baryons. Both properties are sufficiently
distinctive so as not to be confused with inflation.
This makes the task of distinguishing them simpler than in the  general
case \cite{ourpaper} and renders them testable by the current generation
of CMB experiments.

\subsection{Scaling Ansatz for Anisotropic Stress}
\label{ss-anisotropic}

Now let us consider the effect of anisotropic stress
sources that obey the scaling ansatz.  These sources are
represented by 
the basis of Eq.~(\ref{eqn:basispi}),
\begin{equation}
4\pi G a^2 \pi_s = 
\tau^{-1/2} f_B = 
\tau^{-1/2}
 {6 \over B_2^2 - B_1^2}
              \left[ {\sin(B_1 k\tau) \over (B_1 k\tau)}
              -{\sin(B_2 k\tau) \over (B_2 k\tau)} \right].
\label{eqn:pibasis}
\end{equation}
Anisotropic stress affects the CMB in two ways.  It contributes directly
to the gravitational potential $\Psi$ through Eq.~(\ref{eqn:psipi})
and hence the Sachs-Wolfe effect.  It also acts as a force in
the momentum conservation equation [e.g. Eq.~(\ref{eqn:simplefluid})] 
that moves matter around.  Thus
it generates true density and curvature fluctuations inside
the horizon in the same way as the pressure perturbations.

The form of Eq.~(\ref{eqn:pibasis}) implies that it is a source of
white noise fluctuations in $\Psi$ above the horizon.  Due to the
$(k\tau)^2$ factor, we expect that the formation of acoustic oscillations
by anisotropic stresses is delayed compared with formation by pressure
fluctuations.  This shifts the acoustic features toward smaller scales
and further away from the predictions of inflation.
On the other hand, their relatively late formation implies that the feedback
mechanism from the compensating energy density of the photons at Jeans length
crossing is less important leading to a wider range of possible effects in
the CMB anisotropy spectrum.

Let us consider a few specific examples.  
In Fig.~\ref{fig:pressure}b, we show the time evolution of fluctuations
in the photon-dominated era from an anisotropic stress of the
form in Eq.~(\ref{eqn:pibasis}) with $B_1=1$ and $B_2=0.5$.
As the anisotropic stress source turns on at $k\tau \sim B_1^{-1}$,
it acts as a direct source of potential fluctuations $\Psi$.  It
then begins to move matter around.  This produces significant
density and accompanying curvature perturbations which thereafter
dominate the structure of the gravitational potentials, 
i.e. $\Psi \sim -\Phi$.  The result is an effective temperature
$\Theta_0+\Psi$ that first follows $\Psi$ into
a rarefaction stage. The fluid then turns around to fall into the
growing potential wells of the source.  Thus the qualitative
effect of anisotropic stress on the CMB is the same as isotropic
stress: the feature at Jeans crossing corresponds to a rarefaction
in the effective temperature and is suppressed in comparison to the main
compressional feature due to infall into the potential well of the source.
This expectation is borne out by the full Boltzmann calculation
in Fig.~\ref{fig:cls}.  Because the dynamical effects of
anisotropic stress are highly
suppressed outside the horizon, the main features of the peaks
are shifted toward smaller scales than for the pressure model. 

\begin{figure}
\begin{center}
\leavevmode
\hskip -0.5truecm
\epsfxsize=3.5in \epsfbox{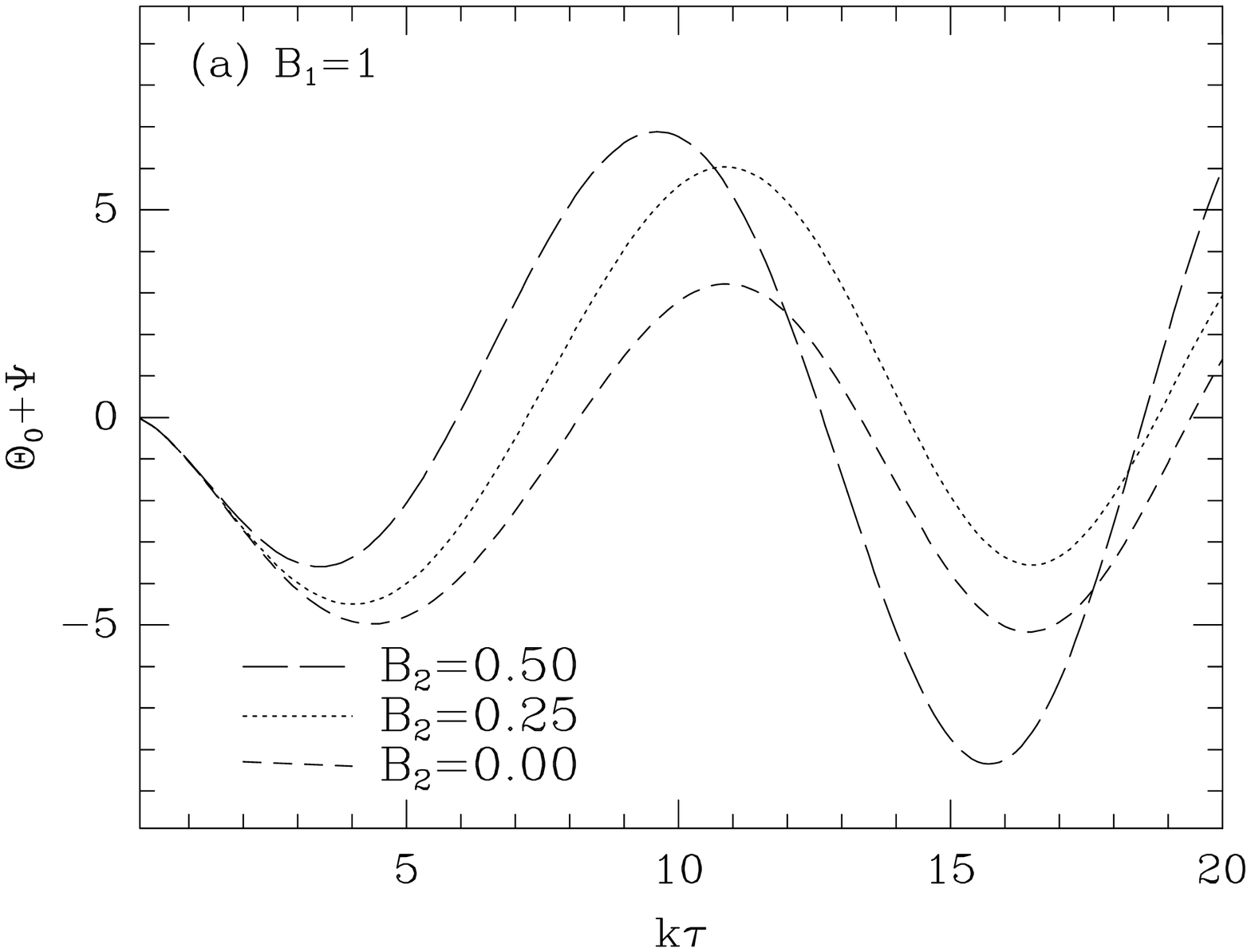}
\epsfxsize=3.5in \epsfbox{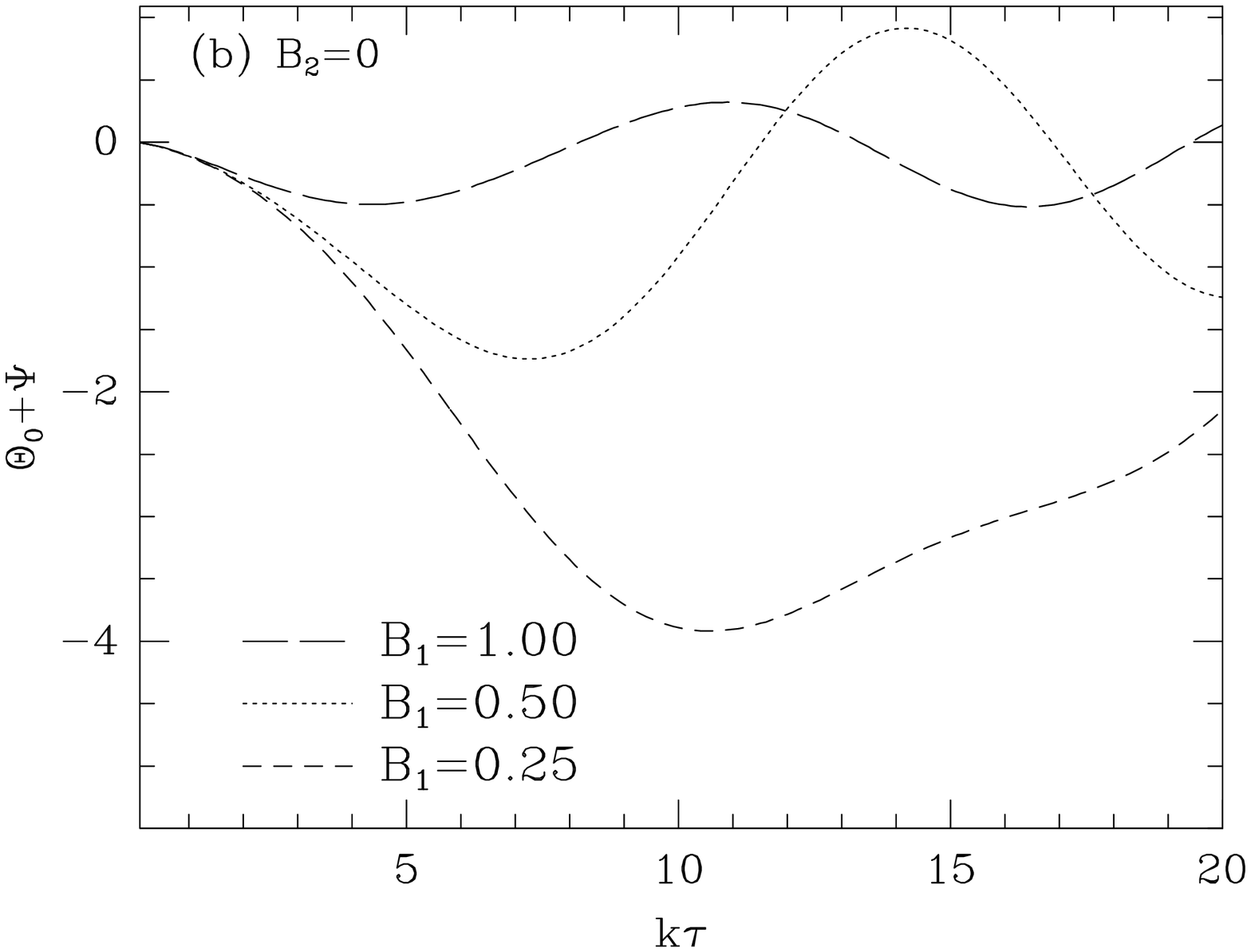} 
\end{center}
\caption{Anisotropic stress scaling source time evolution
(a) $B_2$ controls the decline of $\pi_s$ from its maximum and has little
effect on the acoustic features.
(b) $B_1$ controls the location of the main peak in $\pi_s$ and hence the
location of the main acoustic feature.}
\label{fig:bpar}
\end{figure}

Now let us consider how these results change with the form of
the anisotropic stress source.  The parameters $B_1$ 
and $B_2$ that define the anisotropic stress $\pi_s$ in 
Eq.~(\ref{eqn:pibasis}) control the maximum of $\pi_s$
and the rapidity of its subsequent decline respectively.  Here
we have assumed that $B_1 > B_2$.   In Fig.~\ref{fig:bpar}, 
we show how the time evolution of the effective temperature varies 
with $B_1$ and $B_2$.  Notice that altering the rapidity of the fall 
off through $B_2$ has little effect on the acoustic structure whereas 
decreasing $B_1$ shifts the main features toward later times and 
hence smaller scales.  Thus the anisotropic stress models which
have features closest to those of inflation set $B_1=1$. 

Thus, we can conclude that anisotropic stress fluctuations tend to shift
the main features toward smaller scales.  Like the result for pressure
fluctuations, this implies that any feature that is near the first peak
in an inflationary model must be sub-dominant, leading to a low-high-low
prediction for the heights of the features (see Fig.~\ref{fig:cls}).
As such, these models are easily distinguished from inflation.

\subsection{Underlying Assumptions}
\label{ss-assumptions}

Since the effects of causal pressure and anisotropic stress fluctuations are
both individually distinguishable from those of inflation, one expects
that the combination of the two would result in a spectrum equally
distinguishable unless there is interference between the modes. 
To better quantify our intuition and identify possible loopholes, it is
instructive to recall the physical basis for the differences in the CMB
spectra. 
 
The crucial distinction between all of these isocurvature models and the
inflation is the behavior of the fluctuations during horizon
crossing.  In an isocurvature model, any source density fluctuations at
this epoch must be compensated to keep the total density fluctuation small.
If the photons take part in this compensation, as they must if they are the
dominant dynamical component at the epoch of horizon crossing, this implies
an anticorrelation between the source and photon density fluctuations.
Inflationary models generate adiabatic fluctuations so that the density
fluctuations of all the species are correlated at horizon crossing. 
This leads to observable consequences with the {\it additional} assumption
that overdense regions of the source represent gravitational potential wells.
The Compton drag of the baryons on the photons attempts to compress the
photon-baryon fluid in the potential well.  In inflationary models, the
photons are already overdense inside the well such that this effect enhances
the first peak and subsequently all odd (compressional) peaks.
In isocurvature models, the opposite occurs leading to a reduction of the
first (rarefaction) peak and an enhancement of the second and all even peaks. 

There are two basic assumptions to this chain of reasoning.  The first is that
the photons must play a role in the causal compensation.  It is possible to
construct a model in which the universe is fully matter dominated at horizon
crossing for all observable peaks where this
assumption is invalid \cite{ourpaper}.  However, we have shown in 
\S \ref{ss-pressure} that this does not occur
in a model with the standard thermal history and reasonable cosmological
parameters.  The second assumption is that overdense regions of the source,
here taken to mean all contributions external to the photon-baryon system, 
represent potential wells.  This is generally a reasonable assumption even
in the isocurvature case since the ability of the photon-density perturbation
to counteract the source is diminished as the fluctuation passes the Jeans
length.  Thus source overdensities represent total overdensities.
The Poisson equation implies that overdensities represent positive curvature
fluctuations and hence potential wells {\it if} the anisotropic stress is
negligible in comparison to the density fluctuation 
[see Eq.~(\ref{eqn:psipi})].

The latter assumption opens up the possibility that anisotropic stress
provides a loophole to these arguments.  More specifically, if
$\pi_s > -\rho_s/2$, then {\it under}dense regions of the source represent
potential wells and the above expectation for the relative heights of the
peaks is inverted.  However, this does not occur if we just simply take a
model with large anisotropic stress $\pi_s$ (see \S \ref{ss-anisotropic}).  
The reason is that a large
anisotropic stress moves matter around to create a correspondingly large
density perturbation.  The energy-momentum conservation laws for a seed source 
[from Eq.~(\ref{eqn:conserve}) for wavelengths well below the background
curvature scale] 
\begin{equation}
\begin{array}{rcl}
\dot \rhos + 3{\displaystyle{\dot a \over a}}(\rhos + \ps) \eal - k\vs, \\
\dot \vs + 4{\displaystyle{\dot a \over a}}\vs \eal k\ps -
	\fract{2}{3} k\pis,
\end{array}
\label{eqn:sconserve}
\end{equation}
imply that for $k\tau \gg 1$, $\pi_s$ is typically a strong source of 
density fluctuations.  

Can we ever construct a model in which $\pi_s \simgt -\rho_s/2$?  
An exception to the above arguments occurs if $p_s=2\pi_s/3$. 
The stresses are then so balanced as to maintain a large anisotropic 
stress without generating a correspondingly large density perturbation
(see discussion of Fig.~\ref{fig:pressure}).  
It is {\it not} sufficient to 
have merely $\pi_s = {\cal O}(p_s)$ to achieve this balance.  

For this one exceptional case, 
it is possible here to have gravitational potentials generated
by anisotropic stress instead of density perturbations.  If such a model
additionally has the peaks in the inflationary positions {\it and}
yields approximately constant gravitational potential perturbations,
it is possible
to evade the arguments of Hu \& White \cite{ourpaper}.
Even though the individual effects of pressure and anisotropic stress
fluctuations lead to predictions in accord with the canonical isocurvature
model, the two may cancel in this way to evade such expectations.
Following Turok \cite{turoknew}, we explicitly construct such 
an example in Appendix \ref{sec-mimic}.
Such models rely on a special relation between the pressure, 
anisotropic stress and density fluctuations and are thus unstable
to perturbations in the equation of state [see discussion surrounding
Eq.~(\ref{eqn:unstable})].

In summary, the two assumptions underlying the case for the
distinguishability of inflation from isocurvature models from the
acoustic signature are that the photons are dynamically significant
at Jeans crossing and that potential wells represent overdense regions
in space.  These criteria are satisfied by a wide range of models 
including all those currently under consideration involving defects
which have observable acoustic signatures.

\section{Discussion}
\label{sec-discussion}

All of the causal models for the formation of large scale structure currently
being considered can be divided into two classes:
(a) inflationary models, which have curvature fluctuations on superhorizon
scales and
(b) scaling seeded-models, such as strings and textures.
In the latter case, there are no initial curvature fluctuations and stress
fluctuations only generate them through the causal redistribution of matter
under energy-momentum conservation \cite{Bar,KodSas}.

We have presented a thorough discussion of this process that
can be used to study the
general properties of any model that proposes a causal mechanism for large
scale structure formation without postulating an inflationary epoch.   
We apply these techniques to study a  representative class of scaling models
inspired by Turok \cite{turokletter}.  For models dominated by
white noise isotropic stress fluctuations, the acoustic signature in
the CMB angular power spectrum follows the
canonical signature of a baryon-isocurvature model. 
Physically, this robust
signature arises from the ability of photon-backreaction to drive
the acoustic oscillation \cite{ourpaper}, a feature that must
be included in a self-consistent calculation.
Models dominated by anisotropic stress fluctuations tend to be even
more extreme, with main features pushed toward smaller scales.
Hence both classes are easily
distinguished from inflation by experiments currently
underway (see Fig.~\ref{fig:family}).  
 
A realistic model, such as strings or textures, may contain additional
complications beyond the simple toy-models explored in this paper, such
as tensor and vector contributions as well as 
reionization.  It can also require non-linear evolution of
the sources that couple the normal modes discussed here and 
leave non-gaussian
signatures in the CMB and/or cause decoherence in the oscillation 
\cite{Magetal}.  
However, such complications are likely to make
alternate models less, rather than more, like inflation.

The analysis in this paper also reinforces the conclusions of \cite{ourpaper}:
in an inflationary model, even peaks are produced by rarefaction waves and odd
peaks are produced by compression waves. On the other hand 
in isocurvature models, even
peaks are produced by compression waves and odd peaks are produced by
rarefaction waves.  As long as the energy density
in radiation at decoupling is significant and gravitational 
potential wells represent overdense regions, such a model
cannot reproduce the inflationary CMB signature without the 
equivalent of putting in the features by hand.

\bigskip

\noindent{\it Acknowledgements:}
We thank 
\href{http://www.sns.ias.edu/$\sim$jnb}{J.N. Bahcall}, 
\href{http://physics7.berkeley.edu/cmbserve/ferreira.html}{P.G. Ferreira} 
and
\href{http://www-astro-theory.fnal.gov/Personal/stebbins/welcome.html}{A. Stebbins} 
for useful
discussions, as well as 
\href{http://www.astro.ubc.ca/people/scott/cmb.html}{D. Scott}, 
N.G. Turok, and J. Magueijo
for comments on a draft of this work.
W.H.~was supported by the NSF and WM Keck Foundation.

\vfill
\eject
\appendix

\section{Relativistic Perturbation Theory with Seeds}
\label{sec-perturbation}

We review the formalism of relativistic perturbation theory in this appendix
and derive the results used in the main text.  We start with a discussion of
gauge transformations in general relativistic perturbation theory and then
derive the metric evolution and fluid equations in three commonly employed gauges (see also \cite{VeeSte,Dur}).
Each of these gauges has advantages for particular problems we consider
in the text, and we present all
of the details necessary to transform from one to the other.

\subsection{Gauge Transformations}
\label{ss-gauge}

The most general form of a metric perturbed by scalar
fluctuations is \cite{Bar,KodSas}
\begin{equation}
\begin{array}{rcl}
g_{00} \eal -a^2[1+2A^GQ], \vertsp\nonumber\\
g_{0j} \eal -a^2 B^GQ_j,  \vertsp\nonumber\\
g_{ij} \eal a^2 [\gamma_{ij} + 
	2H_L^G Q \gamma_{ij} + 2H_T^G Q_{ij}] \vertsp,
\end{array}
\label{eqn:metric}
\end{equation}
where $Q$ is the $k$th eigenfunction of the Laplacian, 
{\it i.e.}~$\exp(i{\bf k}\cdot {\bf x})$ in a flat space, 
$Q_i \equiv -k^{-1} Q_{|i}$ and $Q_{ij} = k^{-2} Q_{|ij} + \gamma_{ij}Q/3$
where $|$ denotes a covariant derivative with respect to the background
3-metric $\gamma_{ij}$ of constant curvature 
$K = -H_0^2(1-\Omega_0-\Omega_\Lambda)$.  
The superscript $G$ is employed to remind the reader that the actual values
vary from gauge to gauge.

A gauge transformation is a change in the correspondence between the
perturbation and the background  represented by the coordinate shifts
\begin{equation}
\begin{array}{rcl}
\tilde \tau \eal \tau + TQ, \nonumber\\
\tilde x^i \eal x^i + LQ^i,
\end{array}
\label{eqn:shift}
\end{equation}
where the conformal time $\tau$ is defined through 
$d\tau=dt/a(t)$ with $a$ as the scale factor.
$T$ corresponds to a choice in time slicing and $L$ a choice of 
spatial coordinates.  Under the condition that 
metric distances be invariant,
they transform the metric as \cite{KodSas}
\begin{equation}
\begin{array}{rcl}
A^{\tilde G} \eal A^G - \dot T - \dotaa T, \nonumber\\

B^{\tilde G} \eal B^G + \dot L + kT, \nonumber\\
H_L^{\tilde G} \eal H_L^G - \fract{k}{3}L - \dotaa T, \nonumber\\
H_T^{\tilde G} \eal  H_T^G + kL.
\end{array}
\label{eqn:metrictrans}
\end{equation}
The normal mode decomposition of the scalar part of the stress-energy tensor
for a fluid ($f$) plus seed source ($s$)
yields 
\begin{equation}
\begin{array}{lcl}
T^0_{\hphantom{0}0} \eal -\rho_f - (\rho_f \delta_f^G + \rhos )Q, \vertsp\\
T^0_{\hphantom{0}i} \eal [(\rho_f + p_f)(v_f^G - B^G) + \vs] Q_i, \vertsp\\
T_0^{\hphantom{i}i} \eal -[(\rho_f + p_f)v_f^G + \vs] Q_i, \vertsp\\
T^i_{\hphantom{i}j} \eal [p_f + (\delta p_f^G + p_s)Q] 
	\delta^i_{\hphantom{i}j} + (p_f\pi_f^G + p_s)Q^i_{\hphantom{i}j} \vertsp.
\end{array}
\label{eqn:stressenergy}
\end{equation}
It is occasionally convenient to break the fluid up into its various particle
components, e.g.~$\rho_f\delta_f\rightarrow\sum_f\rho_f\delta_f=\rho_T\delta_T$,
and we shall preserve generality by writing equations applicable to either the
single- or multi-fluid case.
The gauge transformations act on the fluid quantities as \cite{KodSas}
\begin{equation}
\begin{array}{rcl}
v_f^{\tilde G} \eal v_f^G + \dot L, \nonumber\\
\delta_f^{\tilde G} \eal \delta_f^G + 3(1+w_f)\dotaa T, \nonumber\\
\delta p_f^{\tilde G}  \eal \delta  p_f^G + 3c_{f}^2 \rho_f
(1+w_f)\dotaa T, \nonumber\\
\pi_f^{\tilde G} \eal \pi_f^G,
\end{array}
\label{eqn:fluidtrans}
\end{equation}
whereas for the seed source they only generate second order corrections.
Here $w_f = p_f/\rho_f$ defines the equation of state, 
$c_f^2 = \dot p_f/{\dot \rho_f}$ is the sound speed in the fluid, and
we have used ${\dot \rho_f}/\rho_f=-3(1+w_f)(\dot{a}/a)$.
Notice that the anisotropic stress $\pi_f$ has a truly gauge invariant
meaning, and we shall hereafter drop the superscript $G$ from it.

\subsection{Synchronous Gauge}
\label{ss-synchronous}

Let us derive the energy-momentum conservation and Einstein-Poisson equations
in the familiar synchronous gauge and use the gauge transformation above
to relate them to alternate representations.  
The synchronous gauge is defined by $A^S=B^S=0$ implying that proper time
corresponds with coordinate time and that constant spatial coordinates are
orthogonal to constant time hypersurfaces, a natural coordinate system for
freely-falling observers.  From any other coordinate system, it is reached
by the transformation
\begin{equation}
\begin{array}{rcl}
T \eal a^{-1} \int d\tau\ a A^G + c_1 a^{-1}, \vertsp\\
L \eal -\int d\tau\ (B^G + kT^G) + c_2, 
\end{array}
\label{eqn:synchshift}
\end{equation}
where the presence of the integration constants $c_1$ and $c_2$ reflects the
fact that the synchronous condition does not uniquely fix the coordinates.
In the past, this fact has lead to much confusion since coordinate ambiguity
in $T$ appears as a fictitious gauge mode in the density evolution.
It is conventional to define
\begin{equation}
\begin{array}{rcl}
h \eal 6H_L^S, \\
\eta \eal -H_L^S - \fract{1}{3}H_T^S,
\end{array}
\label{eqn:synchmetric}
\end{equation}
as the fundamental metric variables.  
Covariant conservation of the stress-energy contributions of the fluid yields
the continuity equation for the background
$\dot \rho_f = -3(\rho_f + p_f) (\dot a / a)$ and for the perturbations
\begin{equation}
\begin{array}{rcl}
\fract{d}{d\tau} \left(\fract{\delta_f^S}{1+w_f}\right) \eal 
	-(kv_f^S + \dot h/2) - 3 \dotaa \fract{w_f}{1+w_f} \Gamma_f,
\end{array}
\label{eqn:synchcont}
\end{equation}
as well as the Euler equation 
\begin{equation}
\begin{array}{rcl}
\dot v_f^S + \dotaa (1-3c_f^2)v_f^S = 
\fract{c^2_f}{1 + w_f} k\delta_f^S + \fract{w_f}{1 + w_f}k\Gamma_f
- \fract{2}{3} \fract{w_f}{1+w_f} (1-3K/k^2)k\pi_f . 
\end{array}
\label{eqn:syncheuler}
\end{equation}
Here, the non-adiabatic pressure perturbation or ``entropy'' fluctuation
is defined as
\begin{equation}
p_f \Gamma_f = \delta p_f^G - c_f^2 \delta \rho_f^G,
\end{equation}
and is manifestly gauge invariant [Eq.~(\ref{eqn:fluidtrans})].
Likewise, conservation of the seed source gives the equations
\begin{equation}
\begin{array}{rcl}
\dot \rhos + 3{\displaystyle{\dot a \over a}}(\rhos + \ps) \eal - k\vs, \\
\dot \vs + 4{\displaystyle{\dot a \over a}}\vs \eal k\ps -
	\fract{2}{3} k(1-3K/k^2)\pis,
\end{array}
\label{eqn:conserve}
\end{equation}
which are also manifestly gauge invariant.

In this gauge, the Einstein equations are straightforward to derive.
The evolution of the scale factor is determined by
\begin{equation}
\left(\dotaa\right)^2 + K = \frac{8\pi G}{3} a^2 \rho_T,
\end{equation}
and
the metric perturbations are given in terms of the
matter sources as\footnote{In \cite{WhiSco} 
there is a typographical error in Eq.~(A16) and the first of Eq.~(A45). These equations are missing a minus sign. No results are changed.} 
\begin{equation}
\begin{array}{rcl}
(k^2 - 3K)\eta - \dotaa \fract{\dot h}{2} \eal 
	-4\pi G a^2 [\delta_T^S \rho_T  + \rhos], \\
k\dot\eta - \fract{K}{2k} (\dot h + 6{\dot \eta}) 
	\eal 4\pi G a^2 [(\rho_T+ p_T)v_T^S 
	+ \vs], \\
\ddot h + \dotaa \dot h
	\eal -8\pi G a^2 [\delta_T^S \rho_T + 3\delta p_T^S
	+ \rhos + 3p_s ], \\
\ddot h + 6\ddot \eta + 2\dotaa (\dot h + 6\dot\eta) - 2k^2\eta \eal 
	-16\pi G a^2 [p_T \pi_T + \pis].
\end{array}
\label{eqn:synchpoisson}
\end{equation}
Notice that the third equation implies that $h$, unlike $\eta$, is
dependent only on $\rho+3p$.

\subsection{Newtonian Gauge}
\label{ss-newtonian}

The Newtonian gauge is defined by the sheer free condition 
$B^N = H_T^N=0$, and it is conventional to call the remaining 
metric variables the Newtonian potential $\Psi \equiv A^N$ and curvature 
fluctuation $\Phi \equiv H_L^N$.  
{}From an arbitrary gauge, it is reached by the transformation
\begin{equation}
\begin{array}{lcl}
T = -B^G/k + \dot H_T^G/k^2 \quad &[=& -{1 \over 2}(\dot h + 6 \dot \eta)/k^2]   , \vertsp \nonumber\\
L = - H_T^G/k \quad &[=& {1 \over 2} (h + 6\eta)/k ] \vertsp,
\end{array}
\label{eqn:newtshift}
\end{equation}
where we have also specialized it to synchronous gauge in the square brackets.
Thus the Newtonian metric perturbations can be written in terms of their
synchronous counterparts as 
\begin{equation}
\begin{array}{rcl}
\Psi \eal \fract{1}{2} [\ddot{h}+6\ddot\eta+\dotaa(\dot{h}+6\dot\eta)]/k^2,
	\nonumber \\
\Phi \eal -\eta + \fract{1}{2} \dotaa (\dot{h}+6\dot\eta)/k^2 , 
\end{array}
\label{eqn:metricsynchnewt}
\end{equation}
and likewise for the fluid variables
\begin{equation}
\begin{array}{rcl}
\delta_f^N \eal \delta_f^S - \frac{3}{2}(1+w_f)\dotaa 
(\dot h + 6\dot\eta)/k^2, \nonumber\\
\delta p_f^N \eal \delta p_f^S - \frac{3}{2} c_{f}^2 \rho_f(1+w_f)
\dotaa (\dot h + 6\dot\eta)/k^2, \nonumber\\
v_f^N \eal v_f^S + \frac{1}{2}(\dot h + 6\dot\eta)/k \vertsp.
\end{array}
\label{eqn:fluidsynchnewt}
\end{equation}
It is a straightforward exercise in algebra to transform the
synchronous gauge equations.  The conservation equations become 
\begin{equation}
\begin{array}{rcl}
\fract{d}{d\tau} \left(\fract{\delta_f^N}{1+w_f}\right) \eal 
	-(kv_f^N + 3\dot\Phi) - 3 \dotaa \fract{w_f}{1+w_f} \Gamma_f,\\
\dot v_f^N + \dotaa (1-3c_{f}^2) v_f^N \eal
\fract{c^2_f}{1 + w_f} k\delta_f + \fract{w_f}{1 + w_f}k\Gamma_f
- \fract{2}{3} \fract{w_f}{1+w_f} (1-3K/k^2)k\pi_f + k\Psi,
\end{array}
\label{eqn:conservenewt}
\end{equation}
and Einstein-Poisson equations become,
\begin{equation}
\begin{array}{rcl}
(k^2 - 3K)\Phi \eal
 {4\pi G} a^2 
	\left\{\rho_T\delta_T^N + \rho_s +  3\dotaa [(\rho_T + p_T) v_T^N
	+ \vs] /k\right\}, \vertsp \\
k^2(\Psi + \Phi) \eal -8\pi G a^2 (p_T \pi_T + \pis). \vertsp
\end{array}
\label{eqn:poissonnewt}
\end{equation}
It is also useful to note that the gauge transformation properties
imply that from an arbitrary gauge, the Newtonian potential can
be constructed as
\begin{equation}
(k^2 - 3K)\Phi = 4\pi G a^2 
\left(\delta_T^G \rho_T + 3 \dotaa [(\rho_T + p_T)(v_T^G-B^G) + \vs ] /k
\right),
\label{eqn:poisson}
\end{equation}
which is commonly called the gauge-invariant Poisson equation.

\subsection{Comoving Gauge}
\label{ss-comoving}

The comoving gauge (superscript $T$) 
is defined by the vanishing of the energy
density flux $T^0_{\hphantom{0}i}=0$ and the auxiliary condition $H_T^T=0$.
It is also sometimes called the velocity-orthogonal isotropic gauge, the
total-matter gauge and the rest frame gauge.
For convenience, we denote $\xi \equiv  A^T$ and $\zeta \equiv H_L^T$.  
{}From an arbitrary gauge, it is reached by
\begin{equation}
\begin{array}{lcl}
T = [v_T^G + \vs/(\rho_T + p_T) -B^G]/k \quad
  &[=& \{v_T^S + \vs/(\rho_T + p_T)\}/k ], \vertsp\\
L = -H_T^G/k \quad
  &[=& {1 \over 2}(h+6\eta)/k ],
\end{array}
\label{eqn:shifttmg}
\end{equation}
where we have again also specialized to synchronous gauge and used the notation
$(\rho_T+p_T)v_T=\sum_f(\rho_f+p_f)v_f$ to preserve generality in the
multifluid case.
The gauge transformations imply that
\begin{equation}
\begin{array}{rcl}
\zeta \eal  \Phi - \dotaa [v_T^N + \vs/(\rho_T + p_T)]/k \\
      \eal -\eta - \dotaa [v_T^S + \vs/(\rho_T + p_T)]/k, \\
\end{array}
\label{eqn:metrictmg}
\end{equation}
and the comoving density is defined as
\begin{equation}
\delta_f^T = \delta_f^S + 3(1+w_f){\dot a \over a}
		[v_T^S + \vs/(\rho_T + p_T)]/k.
\label{eqn:deltatmg}
\end{equation}
Note that the rhs is the same
if we employ the Newtonian gauge density and velocity perturbation.
Since the velocity is the same as in the Newtonian gauge $v_f^T = v_f^N$, 
we obtain for
the conservation equations
\begin{equation}
\begin{array}{rcl}
\fract{d}{d\tau} \left(\fract{\delta_f^T}{1+w_f}\right) \eal 
	-(kv_f^T + 3\dot\zeta) - 3 \dotaa \fract{w_f}{1+w_f}\Gamma_f,\\
\dot v_f^T + \dotaa (1-3c_f^2) v_f^T  \eal \fract{c^2_f}{1+w_f} k\delta_f^T
	- 3\dotaa c_T^2 [v_T^T + \vs/(\rho_T + p_T)] \\
&& \qquad + \fract{w_f}{1+w_f}k\Gamma_f
	- \fract{2}{3} \fract{w_f}{1+w_f} (1-3K/k^2)k\pi_f  \\
&& \qquad - \fract{k}{k^2 - 3K} 4\pi G a^2(\rho_T \delta_T^T + \rho_s)
	-{8\pi G a^2}(p_T \pi_T + \pis)/k. \\
\end{array}
\label{eqn:conservetmg}
\end{equation}
The evolution of the metric perturbations can be obtained from the relations
(\ref{eqn:synchpoisson}) and (\ref{eqn:syncheuler}) for the synchronous
Poisson and Euler equations
\begin{equation}
\begin{array}{rcl}
\xi \eal - S/(\rho_T + p_T), \vertsp\\
\dot\zeta \eal \dotaa \xi - K[v_T^T + v_s^T/(\rho_T + p_T)]/k, \vertsp
\end{array}
\label{eqn:zetadot}
\end{equation}
where the fundamental source to metric fluctuations is given by the stress
perturbations
\begin{equation}
\begin{array}{rcl}
S \eal c_T^2 \rho_T \delta_T^T + p_T\Gamma_T + \ps 
	- \frac{2}{3} (1-3K/k^2)(p_T \pi_T + \pis) \vertsp\\
  \eal \delta p^T_T + \ps 
	- \frac{2}{3} (1-3K/k^2)(p_T \pi_T + \pis) \vertsp.
\end{array}
\label{eqn:stresssource}
\end{equation}
The fact that stress perturbations act as the direct source of
comoving curvature is important for the causal arguments we make in
\S \ref{sec-conservation}.

\section{Causal Constraints}
\label{sec-causal}

Once the stress fluctuations are known, the causal evolution of matter and
metric fluctuations is determined by energy momentum conservation and the
Einstein-Poisson equations respectively.  
Thus to impose causality on a model, one must
merely ensure that the initial conditions are causal and enforce causal stress
perturbation behavior.  In this appendix, we shall consider these two issues
in detail.

\subsection{Initial Conditions}
\label{ss-initial}

If the initial conditions could be set up when the metric of the
universe was precisely Friedman-Robertson-Walker, they are
trivial: zero perturbations in all quantities initially, independent
of complications such as gauge.  
Realistically however, we can only start the calculation some 
finite time afterwards when stress fluctuations and consequently
some metric, energy density and momentum density fluctuations have
already formed.  
As is evident from Appendix \ref{sec-perturbation}, 
covariant energy-momentum
conservation, and hence the causal constraint on these quantities, 
takes on different
forms in different gauges.  It is useful to pick a representation that
corresponds to our naive intuition for causal evolution discussed in \S \ref{sec-conservation}:
that these three quantities should be negligibly small
near the initial epoch well outside the horizon 
(see also Appendix of \cite{ourpaper}).

We can summarize this intuition as follows: pressure gradients can cause a
change in the momentum density of the matter and hence a bulk velocity of order
$(k\tau)\delta p/(p+\rho)$.
The divergence of the bulk velocity then kinematically forms a density
perturbation of order $(k\tau)^2 \delta p/(p+\rho)$ corresponding to a
curvature fluctuation of order $\delta p/(p+\rho)$ from the Poisson equation.
The fact that this process requires the movement of matter sets the causal
constraint that the energy density fluctuation, momentum density and curvature
fluctuation all must vanish initially.
However, this intuition only holds if metric terms in the energy-momentum 
conservation equations leave the basic form of the conservation equations
unaltered. 

The comoving gauge provides the desired representation.
If we assume that the wavelength of the fluctuation is much less than the
curvature scale of the background, as we shall throughout this section, 
Eq.~(\ref{eqn:zetadot}) implies that the curvature perturbation $\zeta$ in
this gauge only changes under the influence of stress sources, exactly as we
would naively expect.
In the absence of stress sources $\dot\zeta=0$ and the continuity equation of
Eq.~(\ref{eqn:conservetmg}) reduces to an ordinary conservation law for number
density fluctuations in a fluid:
$(\delta n_f^T/n_f)\propto \delta_f^T/(1+w_f)$ and
$d(\delta n_f^T/ n_f)/d\tau = -kv_f^T$.
Furthermore in this gauge we can rewrite Eq.~(\ref{eqn:conservetmg}) purely in
terms of the stresses [using Eq.~(\ref{eqn:zetadot})], making manifest the
intuition developed earlier regarding stresses as the generators of velocities
and thus density perturbations.
The continuity equation for the combined fluid and source components becomes
\begin{equation}
\fract{d}{d\tau}\left( \fract{\delta_T^T \rho_T + \rho_s}{\rho_T + p_T}
	\right)
= -k[v_T^T + v_s/(\rho_T + p_T)] + 3 \dotaa F,
\label{eqn:totalcontinuitytmg}
\end{equation}
where
\begin{equation}
(\rho_T+p_T) F = c_T^2 [\delta_T^T \rho_T + \rhos] - \fract{2}{3}
	(p_T\pi_T + \pis).
\end{equation}

Note that the adiabatic pressure term $\propto c_T^2$ is proportional
to the density perturbations and is thus initially ineffective. 
Furthermore, anisotropic terms are generically suppressed by $k^2$
compared with pressure terms outside the horizon.  
Thus Eq.~(\ref{eqn:totalcontinuitytmg}) implies that energy density
perturbations in this gauge are built up initially by energy density flows.
Combined with the stress sources from the Euler equation, this implies that
$\delta_T^T\rho_T + \rho_s$ will build a tail that scales as
$(k\tau)^2 (\delta p_T^T + \ps)$ for $k\tau \ll 1$ as expected.
Thus our intuition as to the nature of the causal constraint can be carried
directly over to the comoving gauge, unlike the synchronous and Newtonian
gauges where the total density fluctuation is 
not suppressed by $(k\tau)^2$ with 
respect to the pressure fluctuation outside the horizon.  

In comoving gauge, the causal constraint is imposed by assuming that the
curvature $\zeta = 0$ initially.
The above arguments also show that setting the comoving total density to
zero initially is essentially equivalent though slightly more restrictive
as we shall show below. 
For calculational purposes, it is convenient to represent this 
constraint in other gauges.  Recall that the $\zeta$ curvature is
constructed from synchronous gauge perturbations as 
\begin{equation}
\begin{array}{rcl}
k\zeta = - k\eta -\dotaa [v_T^S + v_s/(\rho_T + p_T)],
\label{eqn:zetasynch}
\end{array}
\end{equation}
where as stated above we ignore factors of $K/k^2$ throughout this section.  
To shed more light on this condition, it is useful to recall how energy flux
generates metric perturbations in this gauge
[see Eq.~(\ref{eqn:synchpoisson})],
\begin{equation}
k\dot\eta = 4\pi G a^2 [(\rho_T+ p_T)v_T^S 
	+ \vs].
\label{eqn:doteta}
\end{equation}
Notice if we make the assignment
\begin{equation}
\begin{array}{rcl}
\tau_{00} & \! \equiv \! & -(k^2\eta/4\pi G)Q = [\delta_T^S
	a^2\rho_T + a^2\rhos - 
	\fract{1}{8\pi G } \dotaa \dot h] Q, \\
\tau_{0i} & \! \equiv \! & a^2[(\rho_T + p_T)v_T^S + \vs](i\hat k_i) Q,
\end{array}
\label{eqn:pseudose}
\end{equation}
Eq.~(\ref{eqn:doteta}) takes on the form of a conservation equation
\begin{equation}
\dot \tau_{00} = \partial^i \tau_{0i}.
\end{equation}
In the literature, this quantity is called the stress-energy pseudo-tensor
\cite{VeeSte}
and is ordinarily rather than covariantly conserved.  It is easy to 
see from Eq.~(\ref{eqn:zetasynch}) that the vanishing of the $\zeta$ 
curvature initially
is equivalent to the statement that $\tau_{00}=0$ and $\tau_{0i}=0$,
i.e.~that the pseudo-energy perturbation and the pseudo-momentum-density
vanish initially as one would expect for conserved quantities.
Thus the two sets of initial conditions are entirely equivalent.  

Finally, let us consider the initial conditions for the Newtonian
gauge.  From Eq.~(\ref{eqn:poisson}), 
the Newtonian curvature  $\Phi$ is algebraically related to the
comoving gauge densities as  
\begin{equation}
(k^2 - 3K)\Phi =
 {4\pi G} a^2 (\rho_T\delta_T^T + \rho_s).
\label{eqn:poissontmg}
\end{equation}
This implies that the isocurvature condition in the comoving gauge
$\zeta=0$ should be directly related to the isocurvature condition
in Newtonian gauge (see also discussion in \cite{ourpaper}).
Let us rewrite Eq.~(\ref{eqn:metrictmg}) using the Newtonian continuity
equation (\ref{eqn:conservenewt}) and the derivative of the Poisson
equation (\ref{eqn:poissonnewt}) as \cite{Lyt}
\begin{equation}
\zeta=  \Phi + \fract{2}{3} \fract{1}{1+w_T}
		\left(\fract{a}{\dot a}\dot\Phi - \Psi\right) .
\end{equation}
If anisotropic stress vanishes initially, the equation of state of
the background is constant and $\Phi$ evolves as a power law then
$\zeta \propto \Phi$.  The two curvature fluctuations are 
comparable except in the degenerate case where $\zeta=0$ and
\begin{equation}
\fract{\dot \Phi}{\Phi} = - \dotaa [ \fract{3}{2}(1+w_T)+1 ].
\label{eqn:decaying}
\end{equation}
In the radiation dominated era, $w_T=1/3$ and this represents a mode
that decays as $\Phi \propto \tau^{-3}$.  Thus the condition $\zeta=0$
is equivalent to $\Phi=0$ except for a decaying mode which becomes
negligible well before horizon crossing.  In the Newtonian gauge,
one can thus take $\Phi=0$ or equivalently its source 
$\delta_T^T \rho_T + \rhos =0$
as the initial condition.

\subsection{Stress Structure}
\label{ss-stress}

Causality constrains the possible forms which the stress perturbations
can take.  We generically expect white noise perturbations except
in cases where conservation laws forbid their generation.  In the
latter case, stress fluctuations outside the horizon can fall off
much steeper than white noise. 

Let us first examine the case of a scalar field since it is 
relevant to cosmological defect models.
Generically, the dynamics of a scalar field $\phi$, is governed
by its Lagrangian ${\cal L}(\phi,\dot\phi)$.  The stress-energy tensor
of the scalar field is
\begin{equation}
T_{\mu \nu} = \partial_\mu \phi \partial_\nu \phi - g_{\mu \nu} {\cal L}.
\end{equation}
Thus, if we decompose its stresses, only the isotropic stress or 
``pressure,''
\begin{equation}
p_s = {T_{ii} \over 3} ,
\end{equation}
depends on both $\dot \phi$ and $\vec \grad \phi$, while the anistropic stresses
depend only on $\vec \grad \phi$.

The causality constraint can be expressed as a condition
on the auto-correlation function of $\phi$:
\begin{equation}
\left\langle \phi(\vec r,\tau) \phi(0,\tau)\right\rangle
	  = 0 \qquad {\rm for} \qquad r > \tau
\end{equation}
where $r$ is co-moving distance.
If we expand $\phi(\vec r,\tau)$ in terms of harmonic functions,
this constraint implies that $\phi(\vec k,\tau) \propto k^0$ for 
$k\tau \ll 1$: $\phi$ behaves as ``white noise'' outside the horizon.
The causality constraint also limits the spatial behavior
of the derivatives of $\phi$: $\dot \phi(\vec k,\tau)$ must scale
as $k^0$ or as some positive power of $k$ to avoid producing superhorizon
fluctuations and $\vec \grad \phi(\vec k,\tau) = i \vec k  
\phi(\vec k,\tau)$
must scale as $k^1$ due to the constraint on the behavior of $\phi$.
Thus $T_{00}$ and the isotropic stress for the scalar field scales 
as white noise to lowest order, whereas anisotropic stress scales at 
$k^2$ outside the horizon. 

Just as energy-momentum conservation limits the form of the density
fluctuation, additional symmetries can constrain the superhorizon
scale behavior of the stress tensor.
For example in electromagnetism,
charge conservation restricts the spatial stresses
produced by electromagnetism so that the large
scale behavior of the fields implies that all of
the stresses scale as $k^2$ for small $k$.
Charge conservation implies that causal processes can not create
super-horizon correlations in charge or current density,
nor can local monopoles be created.  Summing random
electric (or magnetic) dipoles leads to an electromagnetic
field whose strength declines as $1/L \sim k$ on superhorizon
scales.  Since the electromagnetic stress tensor,
\begin{equation}
T_{ij} = {1 \over 4 \pi}\left[{1 \over 2}(E^2 + B^2)\gamma_{ij} 
- E_i E_j -B_i B_j \right]
\end{equation}
is quadratic in the field strengths, this small $k$ behavior
of $E$ and $B$ implies that both the isotropic and
anisotropic stresses scale
as $k^2$ for small $k$.  In magnetohydrodynamics, the isotropic
component gives the magnetic pressure whereas the anisotropic
part gives the $\vec J \times \vec B$ force in the Euler equation.
The Newtonian version of these calculations can be 
found in \cite{Wasserman}. In this type of model, the specific signature 
in the CMB from white noise pressure contributions of \S \ref{ss-pressure} is
replaced by the more general properties discussed 
in \S \ref{ss-anisotropic}. 

\section{Mimicking Inflation}
\label{sec-mimic}

As discussed in \S \ref{ss-assumptions}, the ability of anisotropic stress to reverse the sign
of gravity opens up the possibility of a loophole to the arguments behind the
distinguishability of inflation from isocurvature models.
For this to occur by the action of seed sources, the isotropic and anisotropic
stresses must be exactly balanced so as to create no density perturbations from
dynamical effects, yet still allow the anisotropic stress to generate
gravitational potential perturbations.  
More specifically, we require $\pi_s = 3p_s/2$ during the epoch when the
acoustic oscillations form and for the scales on which they are
observable.  
The ($p_s,\pi_s)$ basis employed in the main text is not well suited to
discuss this case since the isotropic and anisotropic stresses are assumed
to be independent sources.  Although that basis is natural for work on causal
constraints, we must now search for an alternate representation to explicitly
build a counterexample.  We show here that properties of the model 
introduced by Turok \cite{turoknew} are a direct consequence of 
enforcing these rather special requirements.

\begin{figure}
\begin{center}
\leavevmode
\epsfxsize=3.5in \epsfbox{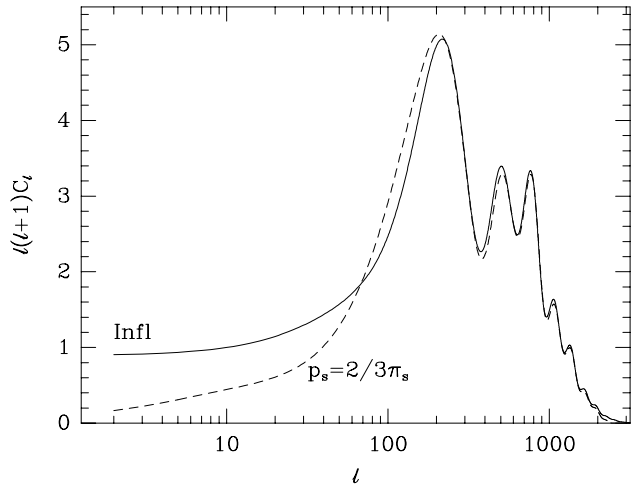} 
\end{center}
\caption{Isocurvature model that mimics inflation.  By choosing
the stress-energy tensor of the seed to reverse the sign of gravity
the general arguments of the main text are evaded.  The anisotropy
spectrum was calculated by a full Boltzmann calculation of the model
of Eq.~(C2) with $A=1$, $B_1=1$, $B_2=0.5$ with cosmological parameters
$\Omega_0=1$, $h=0.5$, $\Omega_b h^2 =0.0125$.}
\label{fig:loophole}
\end{figure}

There are four functions $\rho_s$, $p_s$, $v_s$ and $\pi_s$ that define the
stress-energy tensor of the seed source and two constraint equations from
energy-momentum conservation
[see Eqs.~(\ref{eqn:stressenergy}) and (\ref{eqn:conserve})] 
\begin{equation}
\begin{array}{rcl}
\fract{d}{d\tau} a^2 \rhos + {\displaystyle{\dot a \over a}}a^2 
	(\rhos + 3 \ps) \eal - ka^2\vs, \\
\fract{d}{d\tau} a^2 \vs + 2{\displaystyle{\dot a \over a}} a^2 \vs 
 \eal k a^2 \ps -
	\fract{2}{3} k a^2 \pis,
\end{array}
\label{eqn:cconserve}
\end{equation}
where we assume $K/k^2 \rightarrow 0$.
This leaves two free functions that may be specified.  Since $p_s$ and
$\pi_s$ have different superhorizon scale behavior it is not possible to
apply the desired constraint $\pi_s = 3p_s/2$ directly.
One way to enforce it is to require $a^2 \vs \rightarrow 0$ for $k\tau\gg 1$.
Momentum conservation also implies that $a^2 v_s$ scales as $k$ for
$k\tau\ll 1$.  The remaining condition can be taken as a causal constraint 
on $\rhos+3\ps$.  Note that this choice directly specifies both of the
synchronous gauge gravitational sources [see Eq.~(\ref{eqn:synchpoisson})b,c].

Causality is enforced in the manner of \S \ref{ss-scaling} by requiring 
\cite{turokletter,turoknew}
\begin{equation}
\begin{array}{rcl}
4\pi G a^2(\rho_s + 3p_s) \eal C_1 \tau^{-1/2} 
	\fract{\sin(Ak\tau)}{(Ak\tau)},\\
4\pi G a^2 v_s \eal C_2 \tau^{-1/2} 
	\fract{6}{B_2^2-B_1^2} {1 \over k\tau}\left[
	\fract{\sin{(B_1 k\tau)}}{(B_1 k\tau)} - 
	\fract{\sin{(B_2 k\tau)}}{(B_2 k\tau)}  \right].
\label{eqn:turokbasis}
\end{array}
\end{equation}
For computational convenience, we relax the assumption of pure scaling in
$\rho_s+3p_s$ at the matter-radiation transition, defining
\begin{equation}
C_1 = (\tau \dot a / a)^{-1}, \\
\end{equation}
which requires $C_2$ to take the form,
\begin{equation}
C_2 = -\fract{2}{3} \fract{1}{1+4\tau \dot a /a}.
\label{eqn:prefactors}
\end{equation}
Thus $C_1$ and $C_2$ interpolate between constants in the radiation and
matter dominated epochs.
An examination of Eq.~(\ref{eqn:cconserve}) shows that for $k\tau\gg1$ the
stress-energy components take the form
\begin{equation}
a^2 \rho_s = -3a^2 p_s = -2a^2 \pi_s = {\rm constant}, \qquad a^2 v_s = 0
\label{eqn:turokstressenergy}
\end{equation}
for {\it all} $A$,$B_1$,$B_2$ as desired.  The additional parameters merely
determine at what point the model takes on this special form for the
stress-energy tensor.  Since $\rho_s+2\pi_s$ is the source of gravitational
potential fluctuations of the seed $\Psi_s$, this implies that at late times
\begin{equation}
\Phi_s = {\rm constant}, \qquad \Psi_s = 0,
\end{equation}
and thus overdensities of the seed provide no gravitational attraction
for the other matter components in the universe.  It is important
that $\Phi_s$ and hence $a^2\rho_s$ are constant in order to remove
metric ``stretching'' effects of the source as well as infall.
Again this illustrates the very special nature of 
Eq.~(\ref{eqn:turokstressenergy}): not only must there exist
a relation between the stresses but also some components of $T_{\mu\nu}$
must be {\it constant} while others must be {\it zero}.  We comment
on the stability of this situation below.

By breaking the relation between overdensities (or more strictly speaking
curvature fluctuations) and potential fluctuations the door for mimicking
inflation has been openned.
One still needs to actually {\it reverse} the sign of gravity such that matter
tends to fall out of overdense regions of the seed.  This is readily achieved
if an additional component such as cold dark matter (CDM) 
exists in the universe.
Causality requires that this additional component have density fluctuations
anticorrelated with the seed at horizon crossing.  Since density fluctuations
in this component create gravitational potential wells whereas those in the
seed do not, the net result is that underdense regions of the seed correspond
to gravitational potential wells.  The fundamental criterion for the existence
of a counterexample has now been met (see \S \ref{ss-assumptions}).  
Furthermore,
since these potential wells arise from CDM fluctuations and both
the infall and ``stretching'' gravitational effects of the source
are absent, they are constant in the matter-dominated epoch 
which results in a baryon-drag signal of
alternating peaks that can closely mimic the standard-CDM inflationary
prediction.  
We show an explicit calculation of such a model with $A=1$, $B_1=1$, and
$B_2 =0.5$ in Fig.~\ref{fig:loophole}.  The initial conditions are established
in this synchronous gauge calculation to eliminate the components of the
stress-energy pseudo-tensor by detailed balance of the seed and fluid
components\footnote{Specifically we take $\dot{h}=(12/7)\sqrt{\tau_i}$,
$\delta_\gamma=\delta_\nu=-(4/9)\tau_i\dot{h}$,
$\delta_b=\delta_c=(3/4)\delta_\gamma$ and all other components zero.
Formally the fluid velocities do not vanish, but we found that setting them
to zero initially gave the same answers as including the compensation.}. 

Finally let us briefly discuss the implications of constraints of the form
Eq.~(\ref{eqn:turokbasis}) to support the claim that it is the special form
of the stress-energy tensor in Eq.~(\ref{eqn:turokstressenergy}) rather than
some more general causal property that permits this counterexample.
For example, in the synchronous gauge it might seem that fixing $\rho_s+3p_s$
{\it alone} is sufficient to determine the behavior of CMB fluctuations
\cite{turokletter}.  In synchronous gauge, the metric perturbation is
specified by two functions, $h$ and $\eta$ [see Eq.~(\ref{eqn:synchmetric})].
The important point to note about the metric evolution equations
(\ref{eqn:synchpoisson}) is that $h$, but not $\eta$, is only dependent on
the evolution of $\rho+3p$ type sources.  Likewise the conservation
equations (\ref{eqn:synchcont}) and (\ref{eqn:syncheuler}) imply that before
last scattering, the photon evolution is driven only by $h$.
Thus the synchronous temperature perturbation at last scattering is purely
determined by the assumption for $\rho_s+3p_s$.  This does {\it not} however
imply that the structure of the observed anisotropy is so determined.
To obtain the observed anisotropy, one must free-stream the radiation from the
last-scattering surface to the present.  After last scattering, the
gravitational redshift from $\eta$ which is dependent on the form of $v_s$
generates photon quadrupole fluctuations [see e.g.~\cite{MaBer}~Eq.~(63)].
This is in fact obvious from the Newtonian treatment which mixes $\rho_s$,
$v_s$ and $\pi_s$ in the gravitational source for acoustic oscillations in
the effective temperature [see Eq.~(\ref{eqn:phis})].
Since gauge choice does not affect {\it physical observables}, the two must
predict the same anisotropy for a given source model: they just choose to
divide it into fluid temperature and gravitational redshift in different
manners.  Thus one cannot simply use the fluid temperature in synchronous
gauge to make arguments about the corresponding temperature in the Newtonian
gauge without fully specifying the model. 

Now let us consider whether any other choice besides the asymptotic form of
Eq.~(\ref{eqn:turokstressenergy}) is possible for such seed contributions.
Let us rewrite the conservation equation Eq.~(\ref{eqn:cconserve}) as
\begin{equation}
\left[ {d^2 \over d\tau^2} + 2 \left(\dotaa \right) {d \over d\tau}
        - {1 \over 3}k^2 \right] 4\pi G a^2\rhos
=- \left[ {d \over d\tau} \left(\dotaa\right) + 2\left( \dotaa
        \right)^2 + \fract{1}{3}k^2 \right] 4\pi G a^2(\rho_s + 3\ps) + 
	\fract{8}{3} \pi G a^2 k^2 \pis.
\label{eqn:unstable}
\end{equation}
If we also require $a^2 (\rho_s + 3\ps) \rightarrow 0$ as in
Eq.~(\ref{eqn:turokbasis}), then this equation is dynamically unstable and
requires $a^2\rhos$ to diverge unless $\rho_s=-3p_s=-2\pi_s$.
Thus the only model that can be constructed out of the $\rho_s+3\ps$ form
assumed in Eq.~(\ref{eqn:turokbasis}), or indeed any form which implies
$|1+3p_s/\rho_s|\ll1$ at late times, satisfies
Eq.~(\ref{eqn:turokstressenergy}).
Such models {\it must} have the novel property of anisotropic stress
fluctuations cancelling the gravitational attraction of matter to the
seed overdensities.  Of course, this relation need only hold for
scales upon which acoustic peaks are visible in the CMB.

In summary, the counterexample of Turok \cite{turoknew} reconstructed
here relies on very special
properties of a specific stress-energy tensor and not on general properties of
causally generated fluctuations.

\vskip 2.0 truecm
\noindent{\tt whu@sns.ias.edu}

\noindent{\tt http://www.sns.ias.edu/$\sim$whu}

\vskip 0.5 truecm

\noindent{\tt dns@astro.princeton.edu}

\noindent{\tt http://www.astro.princeton.edu/$\sim$dns}

\vskip 0.5 truecm

\noindent{\tt white@rigoletto.uchicago.edu}

\noindent{\tt http://www-astro-theory.fnal.gov/Personal/mwhite/welcome.html}
\end{document}